\documentclass[sigconf, nonacm]{acmart}
\usepackage{graphicx}
\usepackage{xspace,colortbl,multirow}
\usepackage{amsmath}
\usepackage{cancel}
\usepackage{array}
\usepackage{verbatim}
\usepackage{dsfont}
\usepackage{color}
\usepackage{amssymb}
\usepackage{subfig}
\usepackage{pifont}
\usepackage{multicol}   
\usepackage{hyperref}

\makeatletter
\newif\if@restonecol
\makeatother

\usepackage[lined,boxed,vlined,ruled,linesnumbered]{algorithm2e}
\usepackage{hyperref}

\newtheorem{example}{Example}

\newtheorem{lemma}{Lemma}

\newcommand{\sys}{\ensuremath{{\tt QUEST}}\xspace}

\newcommand{\docs}{\mathcal{D}\xspace}

\newcommand{\attr}{A\xspace}

\newcommand{\totalcost}{\mathcal{C}\xspace}

\newcommand{\wiki}{\ensuremath{\texttt{WikiText}}\xspace}

\newcommand{\swde}{\ensuremath{\texttt{SWDE}}\xspace}

\newcommand{\legal}{\ensuremath{\texttt{LCR}}\xspace}
\newcommand{\pz}{\ensuremath{\texttt{PZ}}\xspace}
\newcommand{\zendb}{\ensuremath{\texttt{ZenDB}}\xspace}
\newcommand{\lotus}{\ensuremath{\texttt{Lotus}}\xspace}
\newcommand{\rag}{\ensuremath{\texttt{RAG}}\xspace}

\newcommand{\nlp}{\ensuremath{\texttt{ClosedIE}}\xspace}
\newcommand{\evaporate}{\ensuremath{\texttt{Eva}}\xspace}

\NewDocumentCommand{\cc}{ mO{} }{\textcolor{red}{\textsuperscript{\textit{CC}}\textsf{\textbf{\small[#1]}}}}
\NewDocumentCommand{\SZZ}{ mO{} }{\textcolor{blue}{\textsuperscript{\textit{SZZ}}\textsf{\textbf{\small[#1]}}}}
\NewDocumentCommand{\Del}{ mO{} }{\textcolor{orange}{\textsuperscript{\textit{Del}}\textsf{\textbf{\small[#1]}}}}

\newcommand{\lei}[1]{\textcolor{purple}{Lei: #1}}

\newcommand{\add}[1]{\textcolor{black}{#1}}

\newcommand{\grammar}[1]{\textcolor{black}{#1}}

\definecolor{shadecolor}{RGB}{220,220,220}

\newcommand\vldbdoi{XX.XX/XXX.XX}
\newcommand\vldbpages{XXX-XXX}
\newcommand\vldbvolume{18}
\newcommand\vldbissue{1}
\newcommand\vldbyear{2025}
\newcommand\vldbauthors{\authors}
\newcommand\vldbtitle{\shorttitle} 
\newcommand\vldbavailabilityurl{https://github.com/qiyandeng/QUEST}
\newcommand\vldbpagestyle{plain} 

\begin{document}

\acmConference[SIGMOD' 25]{Make sure to enter the correct
	conference title from your rights confirmation email}{June 22--27,
	2025}{Berlin, Germany}
\title{QUEST: Query Optimization in Unstructured Document Analysis}

\setcopyright{acmlicensed}
\copyrightyear{2025}
\acmYear{2025}
\acmDOI{XX.XXXX/XXXXXXX.XXXXXXX}
\acmPrice{}
 

\author[]{Zhaoze Sun, Qiyan Deng, Chengliang Chai, Kaisen Jin, Xinyu Guo$^{\dagger}$, \\ Han Han$^{\dagger}$,  Ye Yuan, Guoren Wang, Lei Cao$^{ *\dagger}$}
\affiliation{
    \institution{University of Arizona$^{\dagger}$, MIT$^{*}$, Beijing Institute of Technology}
}


\acmISBN{979-8-4007-0422-2/24/06}

\pagenumbering{roman}
\twocolumn
\pagestyle{plain} 

\renewcommand\thefigure{\Roman{figure}}
\renewcommand\thetable{\Roman{table}}

\clearpage

\pagenumbering{arabic}
\setcounter{page}{1}
\setcounter{figure}{0}
\setcounter{table}{0}
\renewcommand\thefigure{\arabic{figure}}
\renewcommand\thetable{\arabic{table}}

\begin{abstract}
Most recently, researchers \grammar{have started} building large language models (LLMs) powered data systems that allow users to analyze unstructured text documents like working with a database because LLMs are very effective in extracting attributes from documents. In such systems, LLM-based extraction operations constitute the performance bottleneck of query execution due to the high monetary cost and slow LLM inference. Existing systems typically borrow the query optimization principles popular in relational databases to produce query execution plans, which unfortunately are ineffective in minimizing LLM cost. To fill this gap, we propose \sys, which features a bunch of novel optimization strategies for unstructured document analysis.
First, we introduce an index-based strategy to minimize the cost of each extraction operation. With this index, \sys quickly retrieves the text segments relevant to the target attributes and only feeds them to LLMs. Furthermore, we design an evidence-augmented retrieval strategy to reduce the possibility of missing relevant segments. Moreover, we develop an instance-optimized query execution strategy: because the attribute extraction cost could vary significantly document by document, \sys produces different plans for different documents. For each document, \sys produces a plan to minimize the frequency of attribute extraction.
The innovations include LLM cost-aware operator ordering strategies and an optimized join execution approach that transforms joins into filters. Extensive experiments on 3 real-world datasets demonstrate the superiority of \sys, achieving 30\%-6$\times$ \grammar{cost savings} while improving the F1 score by 10\% -27\% compared with state-of-the-art baselines. 

\end{abstract}


\settopmatter{printacmref=false}
\renewcommand\footnotetextcopyrightpermission[1]{}

\maketitle

\section{Introduction}~\label{sec:intro} 
Modern corporations often maintain a large amount of unstructured data including text documents such as web pages. In fact, according to IDC research~\cite{intro_datalake}, unstructured data accounts for 80\%-90\% of the data. 
Recently,  researchers \grammar{have started} building LLMs-powered systems, such as UQE~\cite{dai2024uqe}, ZenDB~\cite{T2T_lin2024zendb}, Lotus~\cite{MMDB_patel2024lotus}, and Palimpzest~\cite{MMDB_liu2024pz}, to analyze the valuable information hidden in text documents. These systems allow a user to select a set of documents, specify some attributes that can be extracted from them, and apply some database analytical operations on these attributes, e.g., filter, aggregation, or join.
The core is to leverage LLMs~\cite{intro_llm1,intro_llm2,intro_llm3} to effectively extract out the attribute values that users are interested in because recent research~\cite{T2T_arora2023evaporate} shows that LLMs are remarkably good at data extraction. These efforts, if successful, can turn unstructured data into actionable insights. For instance, a lawyer may employ this system to swiftly locate legal cases about murder with a minimum of three charges and a 15-year sentence.

Similar to \grammar{traditional} databases, a successful unstructured data analysis system relies on a query optimizer that automatically produces a plan, minimizing query execution costs. However, \grammar{LLMs play a central role in such a system}, raising unique challenges and optimization opportunities. More specifically, compared to traditional database operations, LLM inferences are much more expensive in both execution time and monetary cost~\cite{MMDB_liu2024pz}, no matter whether using commercial LLM services or deploying open source LLMs on high-performance thus expensive GPUs. Because unstructured data analysis relies on LLMs to extract attributes, the extraction operation thus constitutes its performance bottleneck. Therefore, the key optimization objective in this scenario is to minimize the LLM cost incurred during extraction, equivalent to (1) minimizing the LLM cost of each extraction operation, which depends on the number of input tokens to an LLM, and (2) minimizing the frequency of invoking the data extraction operations. 


To achieve the above optimization objective, we propose \sys with two key components: (1) an index-based attribute extraction strategy to minimize the {\it number of input tokens} per extraction; (2) an instance-optimized query execution strategy to minimize the {\it frequency} of attribute extraction.
%
\add{Crucially, these optimization strategies are model-agnostic. Although the cost and performance may vary across different LLMs, the core principles of \sys remain effective in reducing the number of input tokens and extraction frequency.}
\sys thus offers an affordable and scalable approach that allows users to select thousands of documents from a large document collection, specify any attributes out of tens or even hundreds of attributes that these documents potentially contain, and analyze these attributes.   



 \noindent\textbf{(1) Index-based Attribute Extraction.}  \sys designs an index for LLMs to efficiently and effectively extract attributes from documents. Rather than using LLMs \grammar{to scan each document routinely}, it only feeds LLMs the relevant text segments, thus significantly reducing the number of input tokens. This retrieval augmented generation (RAG) inspired solution features two innovative designs: a two-level index and evidence augmented retrieval.

 \noindent\underline{\it Two-level Index.} First, \sys summarizes the subject of each document and indexes this information to filter documents irrelevant to the target attributes. Then \sys builds a segment level index to retrieve relevant segments from the remaining documents. Compared to existing document analysis systems such as ZenDB~\cite{T2T_lin2024zendb}, which filter data only at the segment level, our two-level strategy more accurately identifies the relevant segments because it avoids erroneously recognizing some segments as relevant when they are from documents in different subjects.  
 


 \noindent\underline{\it Evidence Augmented Retrieval.} Second, when searching the index, existing RAG style solutions tend to miss relevant segments, because the query, which typically contains the attributes and their text description, is often not informative enough. To solve this problem, one can rely on users to manually enhance text descriptions like prompt engineering, but it is tedious and time-consuming. We propose a sampling-based approach to automatically collect the {\it evidence} on what the relevant segments should look like and use this evidence to {\it augment} the retrieval.

\noindent\textbf{(2) Instance-optimized Query Execution.} Rather than first extracting out the attribute values involved in a query and then employing the database query optimizer to produce an execution plan, \sys adopts a {\it lazy extraction} strategy. That is, it interleaves attribute extraction and analytical operations (e.g., filter or join), only extracting an attribute when an analytical operation has to evaluate it. In this way, \sys opens optimization opportunities to minimize the frequency of data extraction. For example, if a legal case in a document is not about murder,  \sys does not have to extract other attributes, e.g., the number of charges or the years of the sentence. Intuitively, similar to relational databases, appropriately ordering the filters or the joins in a query could effectively avoid extracting the values that have no chance to appear in the final query results. 

Guided by the optimization objective of \sys, we revisit the query optimization principles in databases, e.g., {\it filter ordering}, {\it predicate pushdown}, and {\it join ordering}. As detailed in Sec.~\ref{sec:optimizer}, rather than simply ordering the filters based on their selectivities and applying the same order to process all documents, \sys produces an {\it instance}-{\it optimized} order for each individual document, leveraging the observation that extracting the same attribute could incur significantly different cost across documents. 

To accurately estimate the extraction cost per document -- the key of this instance optimized strategy, \sys adopts an ``optimize at execution time'' architecture. That is, unlike traditional database optimizers, which produce optimized plans before query execution, \sys estimates the cost during query execution time and produces a plan per document on the fly.

Moreover, we show that pushing down the predicates does not always yield the optimal plan. Instead, \sys introduces a join transformation method that converts a join into a specialized filter and orders it with other filters in the query. We theoretically show that the plan produced in this way is guaranteed to be better than predicate pushdown. For a multi-way join, \sys faces challenges in estimating join selectivities as data records are not available in advance. To address this, we introduce a dynamic join ordering strategy that determines which tables to join during query execution, leveraging our {\it optimize at execution time} architecture.

To summarize, we make the following contributions.

\noindent (1) We propose an \grammar{execution-time optimizer} that produces query plans instance-optimized w.r.t. different documents.

\noindent (2) We propose a join transformation technique that is guaranteed to outperform the classical predicate pushdown strategy.

\noindent (3) We propose a two-level index to effectively filter irrelevant segments, reducing extraction costs without compromising accuracy.

\noindent (4) Extensive experiments demonstrate the superiority of \sys, achieving 0.3--6$\times$ cost savings while improving the F1-score by 10\%-27\% compared with the state-of-the-art baselines.
\vspace{-.2em}
\section{\sys Overview}
\label{sec:overview}

In this section, we first introduce how a user specifies a query and the types of queries that \sys supports. We then describe \sys's overall architecture and highlight the novel designs.


\subsection{User Queries}
\label{sec.user}

In \sys, a user could formulate a query in SQL semantics similar to ZenDB~\cite{T2T_lin2024zendb}. That is, the user selects a subset of documents from her data sources, e.g., the legal documents produced in the last two years. She then specifies some attributes that can potentially be extracted from these documents and applies some analytical operations such as filtering and joining on these attributes. Note that the key techniques of \sys are orthogonal to the query language and thus are compatible with other systems such as Palimpzest~\cite{MMDB_liu2024pz}, which uses Python-style interfaces.  
%
Moreover, users who favor querying in natural language can use \sys, which can employ current NL2SQL techniques~\cite{NL2SQL_ao2024xiyan,NL2SQL_gaotext,NL2SQL_liu2024survey} to convert NL queries into SQL-like queries.



\noindent \textbf{Supported Queries.} 
In this work, we target optimizing Selection-Projection-Join (SPJ) queries over unstructured documents. We leave optimizations on other important types of queries as future work, such as aggregation.

\begin{figure*}[h]
    \centering
    \includegraphics[width=1\linewidth]{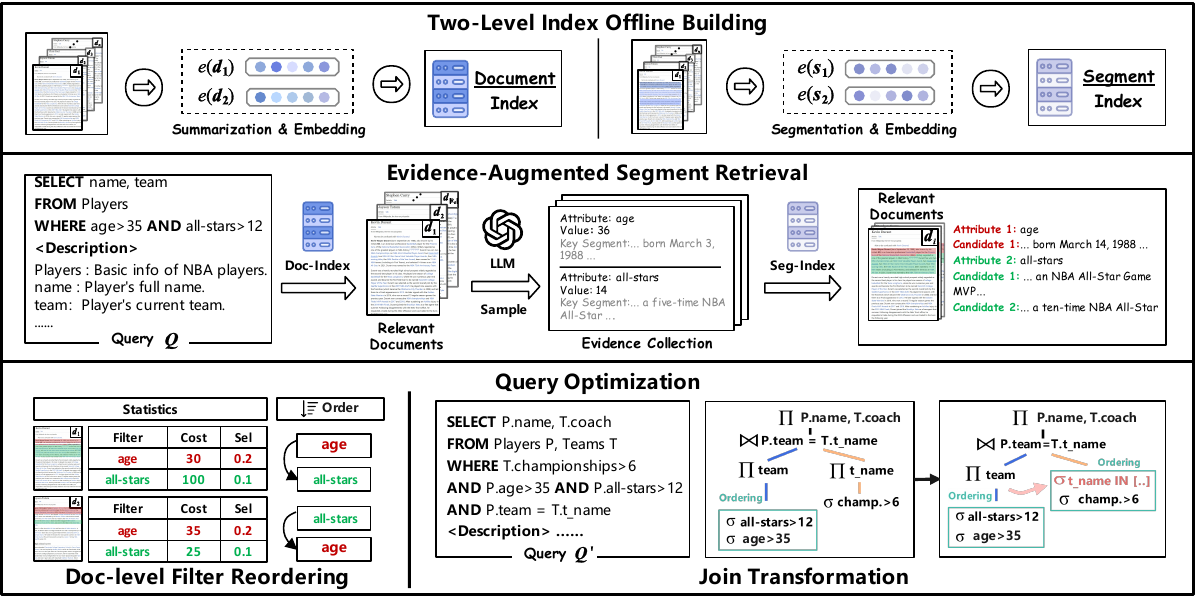}
    \vspace{-2.5em}
    \caption{\sys Framework}
    \vspace{-2em}
    \label{fig:framework}
\end{figure*}

We first introduce some notations. $A$ denotes the set of attributes in the user-specified query $Q$, and $a_i$ denotes each attribute in $A$. 
$\vartheta$ denotes the expression in the \texttt{WHERE} clause of $Q$, consisting of a series of filters, and $\theta_j$ corresponds to each filter in $\vartheta$. \sys supports a broad range of filters. (1) \sys supports queries that are conjunctions or disjunctions of any number of filters.
As an example, in Figure~\ref{fig:framework}, the query $Q$ seeks to find NBA players who are over the age of 35 and have made more than 12 All-Star appearances. The expression $\vartheta$ is a conjunction of these two filters ($e.g.,$ $\vartheta: age > 35\,\texttt{AND}\,all-stars > 12$). 
(2)  Any single filter $\theta_j$
for $a_i\in A$ can be an equality filter ($e.g.,$ $\theta_j: a_i=$"Kevin Durant"), an open range filter ($e.g.,$ $\theta_j: a_i > 27$ ) or a close range filter ($e.g.,$ $\theta_j: 25 \leq a_i \leq 30$).
%
\sys also supports join operations. We use $\mathcal{G}=\{\mathcal{T}, \mathcal{E}\}$ to denote the join graph, where nodes ($\mathcal{T}$) denote the set of tables and edges ($\mathcal{E}$) represent the join relations.



\subsection{Overall Framework}
\label{sec.framework}
Inspired by the principle of relational databases, \sys first indexes the documents and conducts query optimization once receiving a query. However, different from traditional databases, which minimize query execution time, \sys targets reducing LLM costs while ensuring analysis quality. We thus propose several innovative designs to achieve these objectives by (1) minimizing the cost of the attribute extraction via index and (2) minimizing the frequency of calling the extraction via query optimization.


Given a document collection $\docs=\{d_1, d_2, \cdots, d_n\}$, \sys first builds a two-level index over all documents. Then, once the user selects a subset of documents, specifies some attributes in these documents, and composes a query on these attributes, \sys will first {\it sample} some documents, extract the attribute values from them and collect some statistics used by the query optimizer, which aims to minimize the LLM costs of extracting attributes from documents. However, fundamentally different from traditional databases, \sys optimizer does not produce a query plan before query execution. Instead, it uses these statistics and the LLM costs calculated during query execution to produce optimized plans {\it on the fly}. Moreover, given a query, it produces query plans at a {\it document by document} basis instead of presuming a uniform plan for all documents. This is because the attribute extraction cost could vary significantly \grammar{from document to document}. To quickly produce these plans, \sys features novel optimization strategies, which, although lightweight, guarantee to produce optimal plans in many cases. The query execution engine uses index-based data extraction and other analytical operations such as join to execute query plans efficiently.

\sys's key components include: a two-level index-based strategy to minimize the cost of extracting attributes from the documents -- the most fundamental operation in \sys; and (2) a query optimizer that revisits the classical query optimization principles in relational databases, such as filter ordering, predicate pushdown, and join ordering, to avoid unnecessary extraction operations.
%
%
%




\subsection{Index-based Attribute Extraction}
\label{subsec:index-based tuple extraction}
The key idea is to use an index to quickly and accurately identify the text segments that potentially contain the target attributes and only feed these relevant segments to LLMs for extraction. Moreover, we leverage the common patterns shown in the sample segments that have the target attributes to improve the extraction quality. The details are presented in Sec.~\ref{sec:indexing}.

\grammar{At a high level, the index-based tuple extraction resembles Retrieval Augmented Generation (RAG).} That is, \sys divides all documents into segments, encodes each segment into an embedding, and loads the embeddings into a high dimensional vector index such as PQ~\cite{product_quantization} or HNSW~\cite{HNSW}. When a query comes, \sys encodes the name of the relevant attributes and their corresponding text descriptions into an embedding. It then uses the vector index to retrieve the segments with embeddings similar to the query embedding. Then \sys only extracts tuples from these segments. 


\sys optimizes the two key steps of this basic RAG style strategy, namely the {\it indexing} and the {\it retrieval}.

\noindent \textbf{Two-level Index.} In addition to indexing documents at the segment level, \sys constructs a document-level index to filter the documents irrelevant to the to-be-extracted attributes. The segment-level index will only be used to retrieve the relevant segments from the remaining documents, as shown in Figure~\ref{fig:framework}. \sys produces this document-level index by first extracting the {\it key sentences} from each document, encoding these sentences into an embedding $e(d_i)$ to represent the document, and then indexing these embeddings. Next, \sys segments each document, encodes each segment into an embedding $e(s_i)$, and builds the segment-level index.

\noindent\textbf{Evidence-Augmented Segment Retrieval.} Although \sys lets users explain attributes, embeddings from attribute names and descriptions often lack information to find all relevant segments. For example, when \sys extracts the \texttt{age} attribute, simply embedding \texttt{age} alongside its description ``\texttt{Player's age.}'' to search relevant segments may miss the segments such as ``Wardell Stephen Curry II (born March 14, 1988) is an American professional basketball player and point guard ...''. This mismatch occurs because the segment does not explicitly mention ``age'', but instead contains a birthdate and unrelated details like ``point guard''. Therefore, its embedding is not necessarily similar to that of the query.



Rather than relying on users to provide such hints, which in fact is equivalent to a tedious prompt engineering process, \sys automatically collects such valuable information during sampling, serving as {\it evidence} to augment segment retrieval. Specifically, when using LLMs to extract an attribute from document samples, \sys records the segments where a corresponding attribute value is extracted from the sampled documents. Then \sys encodes these segments into representative embeddings and uses each as evidence to retrieve the segments relevant to this attribute from the relevant documents. Finally, it merges all these segments, eliminates duplicates, and inputs them into the LLM for extraction.
\subsection{Query Optimization}
\label{subsec:query optimization}
We revisit the classical query optimization principles in relational databases, such as filter ordering, predicate pushdown, and join ordering, to avoid unnecessary attribute extraction operations. 


\noindent\textbf{Filter Ordering.}
In \sys, the LLM cost of extracting different attributes from a document collection could vary significantly because the \sys index tends to discover various numbers of relevant text segments with respect to different attributes. Similarly, the extraction cost varies significantly document by document, even when extracting the same attribute. This {\it variant cost} observation guides us to propose a new strategy to order filters when multiple filters exist in one query. 

\grammar{First, unlike traditional databases, which simply order the filters based on their selectivities}, \sys orders the filters based on a cost model that takes into consideration both the LLM cost -- the number of tokens of the relevant segments and the selectivities estimated on sample documents. 
In Figure~\ref{fig:framework}, taking query $Q$ as an example, for document $d_1$, the filter on \texttt{age} with a low selectivity but a small number of candidate tokens should have a higher priority than the filter on \texttt{all\_stars} which although has a low selectivity, faces a large number of input tokens, thus incurring high LLM cost. 
Using this new cost model, we design a linear logarithmic time complexity algorithm that produces optimal plans in a broad range of scenarios.

Moreover, because extracting the same attribute from different documents could incur different costs, \sys thoroughly abandons the ``one single order for one query'' methodology in traditional databases and instead produces different orders when evaluating tuples extracted from different documents, as shown in Figure~\ref{fig:framework}.


\begin{example}
\label{example:filter}
For the query in Figure~\ref{fig:framework}, the selectivity of \texttt{all-stars}, which is estimated on some sample documents, is smaller than that of \texttt{age}. However, the LLM costs vary between documents $d_1$ and $d_2$. Therefore, when processing $d_1$, \sys will extract \texttt{age} before \texttt{all-stars}, because the cost of extracting \texttt{all-stars} is significantly higher. On the contrary, the cost of extracting \texttt{all-stars} from $d_2$ is slightly lower than that of extracting \texttt{age}. Thus, \sys will prioritize \texttt{all-stars} over \texttt{age} due to its low selectivity.
\end{example}

\noindent\textbf{Join Transformation.} 
\grammar{Although \sys could execute a join by first extracting the attributes of the two tables involved in the join and then applying an existing join algorithm such as hash join, this is sub-optimal due to the prohibitive attribute extraction cost}, while minimizing the extraction cost is \sys's key objective. Therefore, we propose a join transformation strategy that first extracts the join attribute of one table and then uses the extracted values as filters to filter the other table (as shown in the bottom right of Figure~\ref{fig:framework}). In other words, \sys transforms a join into a filter. Treating this automatically generated filter equally to other filters, the \sys optimizer uses the cost model discussed above to order these filters. In this way, \sys might prioritize joins over filters to minimize the LLM cost, contradictory to the predicate pushdown principle in traditional databases.

\begin{example}
\label{example:join}
In Figure~\ref{fig:framework}, \texttt{Players} and \texttt{Teams} join on the \texttt{team} (\texttt{t\_name}) attribute with the filters "\texttt{P.age}>35 \texttt{AND} \texttt{P.all-stars}>12" and "\texttt{T.championships}>6". Suppose that after applying "\texttt{P.age}>35 \texttt{AND} \texttt{P.all-stars}>12", only a few documents satisfy the filters on \texttt{Players}; and the values of the \texttt{team} attribute only include Warriors, Lakers and Celtics in these remaining documents. \sys will add an \texttt{IN} filter "\texttt{T.t\_team} IN [Warriors, Celtics, Lakers]" to \texttt{Teams} and order it with the existing filter "\texttt{T.championships}>6". If the new filter has low selectivity and cost, it is likely to be executed prior to "\texttt{T.championships}>6", thus saving LLM cost.
\end{example}



\noindent\textbf{Dynamic Join Ordering.}
Different join orders may incur significantly different costs when multiple joins exist in a query. To reduce search space, traditional databases typically use dynamic programming to identify an effective order. However, accurately estimating join selectivities,  \grammar{the key to effectively ordering} joins, is still an open problem. This is even worse in \sys because the attribute values and the tables are not available beforehand. To address this issue, we introduce an algorithm that dynamically and progressively decides the join order during query execution. More specifically, \sys first selects two tables to join based on our cost model, and it will determine the next join only after the first join finishes execution. This process iterates in a left-deep manner until all joins have been executed. In this way, every time \sys decides which table to join, the left table has already been extracted. This effectively alleviates the problem of estimating join selectivity. 

In the remainder of this paper, we first introduce our query optimization techniques to minimize the frequency of attribute extraction (Sec.~\ref{sec:optimizer}), which constitute our core technical novelties. Then, in Sec.~\ref{sec:IndexDefinition}, we introduce our index-based attribute extraction method, which minimizes the LLM cost of each extraction operation while improving the accuracy.

\noindent \textbf{Remark.} \add{Although our primary focus is on optimizing the LLM cost for executing individual query, \sys's architecture naturally supports multiple concurrent queries. First, the two-level index built offline efficiently supports concurrent queries. Furthermore, the instance-optimized nature of query execution, processing documents either individually or in batches, enables straightforward intra-query parallelism: a query spanning numerous documents can be partitioned across multiple workers, each executing the optimized plan for its document subset. Crucially, \sys is able to cache the results of LLM attribute extractions. By storing and reusing previously extracted tuples, redundant extraction operations can be eliminated among concurrent queries, significantly reducing overall cost and latency. These aspects, combined with standard database concurrency control mechanisms, ensure that \sys is effective in serving simultaneous user requests. However, in-depth exploration of specific optimizations and benchmarking the performance of multiple query processing are beyond the scope of this paper and represent important directions for future work.}

\section{QUERY OPTIMIZATION}
\label{sec:optimizer}
\begin{sloppypar}
We first discuss the filter ordering optimization in Sec.~\ref{subsec:filter}. Then, we present the optimizations with respect to join in Sec.~\ref{subsec:join}, including the optimization of one single join and join ordering.

%
%

\subsection{Filter Ordering}
\label{subsec:filter}
As introduced in Section~\ref{subsec:query optimization}, \sys differs from relational databases on filter ordering strategy, because \sys's key objective is to minimize the LLM cost in extracting tuples. In this section, we introduce our filter ordering method that takes both the number of input tokens to LLMs and the selectivities of the filters into consideration, yields optimal orders with respect to different combinations of filters, e.g., conjunction, disjunction, or a mix of both.

\subsubsection{The Filter Ordering Problem in \sys\nopunct}\ \\

The filter ordering in \sys aims to find an order of extracting attributes and evaluating the corresponding filters to minimize the LLM cost. Furthermore, as discussed in Section~\ref{subsec:query optimization}, because the extraction cost could vary significantly document by document, \sys produces different orders with respect to different documents.
Formally, we use $ \mathcal{C}_{Q}(o)$ to denote the total cost of executing query $Q$ on a document with an order (denoted by $o$) of the filters in $\vartheta$. The goal is to find the optimal order $o^*$:
\vspace{-1pt}
 \begin{equation}
 \small
        o^*=\arg\min_{o\in O}\mathcal{C}_{Q}(o)
    \end{equation}

Next, we introduce our filter ordering method and analyze its optimality considering different combinations of filters, namely conjunction, disjunction, and a mix of both.

\subsubsection{Conjunctions\nopunct}\ \\
Consider a query $Q$ that includes a \texttt{WHERE} clause containing a conjunction of filters.
Let $A_s \subseteq \attr, A_w \subseteq \attr$ denote the set of attributes that appear in \texttt{SELECT} and \texttt{WHERE} clauses, respectively. 
For each filter $o[i]$ (i.e., the $i$-th filter in the order), the cost associated with extracting its relevant attribute is represented as $c^F[i]$, and its selectivity is given by $p[i]$.
For each attribute $a_j \in A_s$, the generation cost is denoted by $c_j^E$. 
Therefore, for a given order $o$, the expected query cost can be represented as follows:

\begin{equation}
    \label{eq:conjunctions}
    \small
    \vspace{-.8em}
    \mathcal{C}_{Q}(o)=\sum_{i=1}^{|o|}c^{F}[i]\prod_{j=1}^{i-1}p[j] + \big(\sum_{j=1}^{|A_s|}c_j^{E}\big) \prod_{i=1}^{|o|}p[i]
\end{equation}

\noindent where $\prod_{j=1}^{i-1}p[j]$ represents the likelihood that filter $c^{F}[i]$ has to be processed by an LLM, given that all its preceding filters in $o$ return \texttt{True}.
$\prod_{i=1}^{|o|}p[i]$ represents the probability of all filters returning \texttt{True}. Only in this case, \sys has to extract the attributes in the \texttt{SELECT} clause, leading to a cost of $\sum_{j=1}^{|A_s|}c_j^{E}$.

\noindent \underline{\textit{Remark.}} Note that for conjunctions, $A_s$ will have to be extracted only if all filters return \texttt{True}. Therefore, in this scenario, \sys should always extract the attributes in $A_w$ first, followed by $A_s$. As a result, when \sys computes the optimal order for conjunctions, it only has to consider the cost of extracting attributes in $A_w$, i.e., only the first term in Equation~\ref{eq:conjunctions}.

\noindent \underline{\textit{Optimal Order.}}
To find the optimal order with respect to each document, the brute-force method is to enumerate all possible orders, which has a time complexity of $\mathcal{O}(|D_Q^*|\times |A_w|!)$ and therefore is too costly. Our key insight here is that using the cost model in Eq.~\ref{eq:priority_conjunction}, \sys is able to find the optimal order $P^*$ in linear logarithmic time. 

First, given a filter $\theta_k \in \vartheta$, \sys uses the index to retrieve the segments relevant to the corresponding attribute and estimates its cost $c_k$ which is proportional to the number of tokens in these segments. \grammar{\sys then uses its selectivity $p_k$ estimated on the sampled table, and the cost $c_k$ to compute a priority score, which determines the position of each filter $\theta_k$ in the optimal order.}
\begin{lemma}
    Sorting filters in descending order based on the following priority score minimizes the expected query cost. 
    \begin{equation}
    \label{eq:priority_conjunction}
    \small
    priority(\theta_k)=\frac{1-p_k}{c_k}, \theta_k \in \vartheta
    \end{equation}
\end{lemma}

Intuitively, Eq.~\ref{eq:priority_conjunction} prioritizes a filter that is more likely to return \texttt{False} and has a low LLM cost. The proof of this lemma can be found in our technical report~\cite{full_version}.

\subsubsection{Disjunctions\nopunct}\ \\
\grammar{For disjunctions, a better order should instead prioritize a filter that is more likely to return \texttt{True}.} \sys thus has a better chance to short-circuit other filters. Accordingly, we modify the cost model from Eq.~\ref{eq:conjunctions} to the equation below:
\vspace{-0.7em}
\begin{equation}
    \label{eq:disjunctions}
    \small
    \mathcal{C}_{Q}(o)=\sum_{i=1}^{|o|}c^{F}[i]\prod_{j=1}^{i-1}\left(1-p[j]\right) + \big(\sum_{j=1}^{|A_s|}c_j^{E}\big)\big[1-\prod_{i=1}^{|o|}\left(1-p[i]\right)\big]
\end{equation}

\noindent \underline{\textit{Optimal Order.}} In this scenario, \sys can still produce the optimal order by sorting the filters by their priority scores in descending order if the score is computed slightly different from Eq.~\ref{eq:priority_conjunction}.

\begin{equation}
\label{eq:priority_disjunction}
\small
priority(\theta_k)=\frac{p_k}{c_k},\theta_k \in \vartheta.  
\end{equation}

However, in the disjunction scenario, the attributes $A_s$ in \texttt{SELECT} clause have to be handled differently, especially when an attribute exists in both the \texttt{WHERE} and \texttt{SELECT} clauses, e.g., $A_s \cap A_w \neq \emptyset$. In this scenario, to correctly evaluate the query, these attributes must be extracted for the following reasons: (1) if one filter in $A_w$ returns true, \sys has to extract them as the output of the query; (2) if no filter in $A_w$ returns true, although \sys does not produce any output, it still has to extract these attributes. This is because in such circumstances \sys has to examine all filters in $A_w$, while these attributes are also part of $A_s$.



Therefore, if $A_s \cap A_w \neq \emptyset$, \sys first extracts the attributes in $A_s \cap A_w$, followed by sorting and executing the filters in $A_s \setminus (A_s \cap A_w)$ based on their priority scores, which still guarantees the optimal result. The proof is analogous to the way we prove the optimality of conjunction and, hence, is omitted here.

\subsubsection{Conjunctions and Disjunctions\nopunct}\ \\
The aforementioned method can be extended to queries that involve both conjunctions (\texttt{AND}) and disjunctions  (\texttt{OR}).

\begin{figure}[h]
    \centering
    \includegraphics[width=0.55\linewidth]{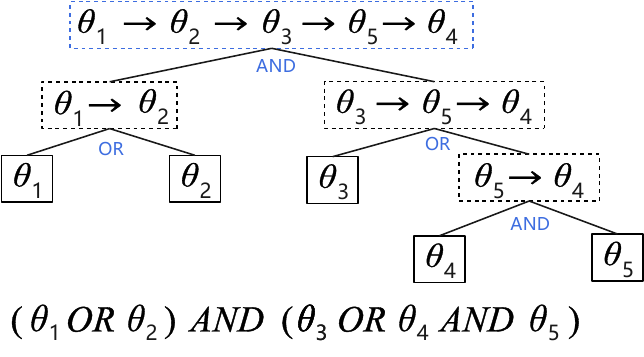}
    \vspace{-1.5em}
    \caption{Example of the Expression Tree.}
    \vspace{-1.3em}
    \label{fig:expression}
\end{figure}

The observation here is that any expressions $\vartheta$ within the \texttt{WHERE} clause is a boolean expression, which can be represented naturally as an expression tree~\cite{expression_tree}; and the execution of the expression is equivalent to a postorder traversal of the tree. As shown in Figure~\ref{fig:expression}, 
each leaf node corresponds to a filter and each non-leaf node denotes a conjunction or disjunction of its children with the same operator precedence. 
Considering Figure~\ref{fig:expression},  $(\theta_1\,\texttt{OR}\,\theta_2)\,\texttt{AND}\,(\theta_3\,\texttt{OR}\,\theta_4\,\texttt{AND}\,\theta_5)$ is the expression within the \texttt{WHERE} clause of $Q$, where the leaf nodes correspond to each filter, i.e., $\theta_1$ to $\theta_5$.
 $\theta_1, \theta_2$ are siblings because they have the same operation precedence.

Representing the expression with a tree structure naturally breaks it down into several sub-expressions. The expected total cost can be computed as a weighted sum of the expected costs of these sub-expressions, with the weights being the probability of each sub-expression evaluated to be \texttt{True}, namely the selectivity. As the weight (selectivity) assigned to each sub-expression is invariant to the order, minimizing the total expression cost is thus equivalent to minimizing the cost of each sub-expression. This property allows us to use dynamic programming to identify the optimal order of these sub-expressions, where the above sorting strategy can be applied to order the filters within each sub-expression. Equation~\ref{eq:dp} formalizes the optimization objective. 

\vspace{-0.3em}
\begin{equation}
    \label{eq:dp}
    \small
    \mathcal{C}^*(\vartheta_T)=
    \begin{cases}
    \displaystyle \underset{o\in O(\vartheta_T)}{\min}\bigg(\overset{|\vartheta_T|}{\underset{i=1}{\sum}}\mathcal{C}^*(o[i])\overset{i-1}{\underset{j=1}{\prod}}p_{o[j]}\bigg),\,\text{for \texttt{AND}}\\
    \displaystyle \underset{o\in O(\vartheta_T)}{\min}\bigg(\overset{|\vartheta_T|}{\underset{i=1}{\sum}}\mathcal{C}^*(o[i])\overset{i-1}{\underset{j=1}{\prod}}(1-p_{o[j]})\bigg),\,\text{for \texttt{OR}}\\
    \displaystyle c_{\vartheta_T},\quad\text{for a single filter}, i.e.,|\vartheta_T|=1
    \end{cases}
\end{equation}
\noindent where $\vartheta_T$ represents the set of children (i.e., sub-expressions) of the current node.
$o \in O(\vartheta_T)$ is one of the possible orders of the sub-expressions in $\vartheta_T$, where $o[i]$ is the $i$-th sub-expression. 
For example, if $T$ is the root of the tree in Figure~\ref{fig:expression}, then $\vartheta_T = \{(\theta_1\,\texttt{OR}\,\theta_2),(\theta_3\,\texttt{OR}\,\theta_4\,\texttt{AND}\,\theta_5)\}$ .
$p_{o[j]}$ denotes the probability of the $j$-th sub-expression in the order being \texttt{True}. Here, $O(\vartheta_T)$ refers to all possible orders of sub-expressions in $\vartheta_T$ and $|O(\vartheta_T)|=2$. 
For the boundary condition, if $\vartheta_T$ is a single filter, $\mathcal{C}^*(\vartheta_T)$ equals the cost of this filter. We implement the algorithm in a recursive manner in Algorithm~\ref{alg:filter}.
The overall time complexity is $O(|\vartheta|log|\vartheta|)$. 

\begin{example}
    For the expression in Figure~\ref{fig:expression}, take the leaf node $\theta_1$ as an example. Based on the boundary condition in Equation~\ref{eq:dp}, its optimal expected cost $C^*(\{\theta_1\})$ equals to the cost of $\theta_1$ itself. Using the cost and selectivity of $\theta_1$, we can calculate its priority score (Line~\ref{alg2:leaf node start}-Line~\ref{alg2:leaf node end}). The same applies to $\theta_2$. Then, employing the first formulation from Equation~\ref{eq:dp}, we calculate the optimal cost $C^*(\{\theta_1, \theta_2\})$ based on the priority score of $\theta_1$ and $\theta_2$, thus resulting in the order $\theta_1 \rightarrow \theta_2$ (Line~\ref{alg2:ordering}). Subsequently, the recursive call returns this order as the optimal order for the sub-expression $\theta_1\,\texttt{OR}\,\theta_2$ (Line~\ref{alg2:sub-expression start}). We can further calculate the cost and selectivity to derive the priority score of $\theta_1\,\texttt{OR}\,\theta_2$ (Line~\ref{alg2:sub-expression start}-Line~\ref{alg2:sub-expression end}). Similarly, we have the optimal order $\theta_5 \rightarrow \theta_4$ for the sub-expression $\theta_4\,\texttt{AND}\,\theta_5$ as well as its priority score (Notably, similar to traditional databases, $\texttt{AND}$ is executed with higher precedence than $\texttt{OR}$ here). \grammar{Next, we sort the sub-expressions $\theta_3$ and $\theta_4\,\texttt{AND}\,\theta_5$ according to the $priority$ of $\theta_3$ and $\theta_4\,\texttt{AND}\,\theta_5$, obtaining $\theta_3 \rightarrow \theta_5 \rightarrow \theta_4$ (Line~\ref{alg2:leaf node start}-Line~\ref{alg2:ordering}).} \grammar{Finally, for the full expression that serves as the entry point for recursion (Line~\ref{alg2:main}), i.e., the root node, we derive the final optimal execution order: $\theta_1 \rightarrow \theta_2 \rightarrow \theta_3 \rightarrow \theta_5 \rightarrow \theta_4$.}
\end{example}

\vspace{-1em}
\begin{algorithm}[h]

    \caption{Filter Ordering}
    \label{alg:filter}
    \KwIn{The expression $\vartheta_T$ within the \texttt{WHERE} clause of $Q$.}
    \KwOut{The optimal order $o^*$.}
    \SetKwFunction{reorder}{Reorder}
    \SetKwFunction{statistics}{Statistics}
    \SetKwFunction{optimalorder}{OptimalOrder}
    \SetKwProg{Fn}{def}{:}{}
    \newcommand{\mycommentstyle}[1]{\color{blue}{\small #1}}
    \SetKwComment{Comment}{\mycommentstyle{  // }}{}
    
    \Fn{\reorder{$\vartheta_T$}}{
        $prior.init()$;\\
        \For{$\vartheta_i$ $\in$ $\vartheta_T$}{
            \If{$|\vartheta_i|=1$}{
                
                $c_i, p_i$ = \statistics{$\vartheta_i$};\Comment{\textcolor{blue}{\small Leaf node}}\nllabel{alg2:leaf node start}
                $prior.append(\vartheta_i, c_i, p_i)$;\nllabel{alg2:leaf node end}
            }
            \Else{
                $o^*_i$ = \reorder{$\vartheta_i$};\Comment{\textcolor{blue}{\small Non-leaf node}}\nllabel{alg2:sub-expression start}
                $\mathcal{C}^*(\vartheta_i), p_i$ = \statistics{$o^*_i$};\\\nllabel{alg2:sub-expression mid}
                $prior.append(o^*_i, \mathcal{C}^*(\vartheta_i), p_i)$;\\\nllabel{alg2:sub-expression end}
            }
        }
        
        $o^*$ = \optimalorder{$prior$};\Comment{\textcolor{blue}{\small$w.r.t.$ Equation~\ref{eq:dp}}}\nllabel{alg2:ordering}
        \KwRet{$o^*$}
    }
    $o^*$ = \reorder{$\vartheta_T$};\\\nllabel{alg2:main}
    \KwRet{$o^*$}
\end{algorithm}
\vspace{-1em}

\subsection{Query Optimization For Join}
\label{subsec:join}

Next, we introduce the optimization techniques with respect to join queries. We begin with optimizing one single join, and then extend our discussion to queries involving multiple joins.


\subsubsection{Single Join: Joining Two Tables\nopunct}\ \\
\label{subsec:single}
Consider a query that joins two tables $T_1$ and $T_2$ on attribute $a$ (i.e.,$ T_1.a = T_2.a'$). The query also has multiple filters $\vartheta$. We abuse $\theta_1$ and $\theta_2$ a little to denote the filters on $T_1$ and $T_2$ respectively.

Typically, relational databases apply filters before joins to reduce the number of tuples in costly join operations. However, to minimize LLM costs, we find that converting joins into filters and optimizing them with other filters is often more effective in our scenario.

\begin{figure}[h]
    \centering
    \includegraphics[width=0.8\linewidth]{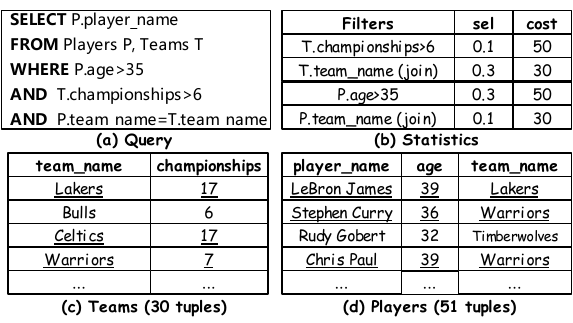}
    \vspace{-1.5em}
    \caption{Example for \texttt{JOIN}}
    \vspace{-0.8em}
    \label{fig:example-join}
\end{figure}

Taking the query in Figure~\ref{fig:example-join} as an example, $\theta_1$ corresponds to the filter on $T_1$ (the \texttt{Teams} table), where $T.championships > 6$, and $\theta_2$ corresponds to the filter on $T_2$ (the \texttt{Players} table), where $P.age > 35$. A traditional query optimizer first pushes down $\theta_1$ and $\theta_2$ to $T_1$ and $T_2$, respectively. It then performs a join on the returned documents (i.e., the tuples underlined in Figure~\ref{fig:example-join}-c). We establish a cost model to compute the expected cost under this optimization.

\noindent [\underline{Plan \ding{172}: \textit{Push $\theta_1$ to $T_1$, $\theta_2$ to $T_2$ and join}}]
The expected cost under the optimal order can be calculated as:
\vspace{-0.7em}
\begin{equation}
\label{eq:172}
\vspace{-0.5em}
\small
Cost(\theta_1(T_1) \bowtie \theta_2(T_2)) = 
\sum_{i=1}^{|T_1|} \totalcost_1^i + p_1\sum_{i=1}^{|T_1|} c_a^i +
\sum_{i=1}^{|T_2|} \totalcost_2^i + p_2\sum_{i=1}^{|T_2|} c_{a'}^i
\end{equation}

\noindent where $\totalcost_1^i$ is the expected cost of executing $\theta_1$ on the $i$-th document of $T_1$. $p_1$ is the likelihood of having to extract $a$ after performing $\theta_1$, and $c_a^i$ is the cost of extracting $a$ from the $i$-th document in $T_1$. The expected cost of $\theta_2(T_2)$ is calculated in the same way.

According to Figure~\ref{fig:example-join}, the cost of executing $\theta_1$ on $T_1$ can be computed as $\scriptstyle\mathit{\sum_{i=1}^{|T_1|} \totalcost_1^i = |T_1| \times c_{1}^i = 30 \times 50 = 1500}$, while the cost of extracting $a$ on $T_1$ can be computed as $\scriptstyle\mathit{p_1\sum_{i=1}^{|T_1|} c_a^i = p_1 \times|T_1| \times c_a^i =0.1 \times 30 \times 30 = 90}$. Similarly, we have $\scriptstyle\mathit{\sum_{i=1}^{|T_2|} \totalcost_2^i = 2550}$ and $\scriptstyle\mathit{p_2\sum_{i=1}^{|T_2|} c_{a'}^i = 459}$ on $T_2$. Finally, we have $\scriptstyle\mathit{Cost(\theta_1(T_1) \bowtie \theta_2(T_2)) = 4599}$.


However, in our scenario, the optimal plan might not be achieved. We analyze the join operation on unstructured documents, consisting of two steps: extracting values of the join attribute from one table and finding matches in the other table. Since the optimization goal of \sys is to minimize data extraction, a join may not be more costly than a filter operation and could have a higher priority. Similar to filters, the priority of a join operation relies on its own data extraction cost and the potential extraction cost it may introduce to the other table, which in turn is determined by the number of matches that each tuple could potentially find, namely the join selectivity. Consequently, the priority of a join can be determined in the same manner as that of a filter. We are now prepared to order operations in a query with one join and multiple filters.



\noindent [\underline{Optimal Plan: \textit{Sort Join and Filters Together.}}]
As shown in Eq.~\ref{eq:join}, the cost of the optimal order corresponds to:

\vspace{-1em}
\begin{equation}
\label{eq:join}
\small
\vspace{-.5em}
Cost^*(\theta_1(T_1) \bowtie \theta_2(T_2)) = 
\sum_{i=1}^{|T_1|} \hat{\totalcost}_1^i +
\sum_{i=1}^{|T_2|} \hat{\totalcost}_2^i
\vspace{-.2em}
\end{equation}
\noindent where $\hat{\totalcost}_1^i$ represents the optimal expected cost obtained by sorting $\theta_1$ with join on the $i$-th document in $T_1$. The same applies to $\hat{\totalcost}_2^i$.


\grammar{Unfortunately, producing this optimal solution requires an accurate estimation of the join selectivity}, which is known to be a notoriously hard problem in databases. It is even worse in our scenario, where the tables are in fact not available beforehand. 

To tackle this challenge, we propose an approach that transforms a join operation into a filter operation and progressively orders the operations during query execution. First, it chooses one table and executes the respective operations, i.e., pushing down the filters on it and then extracting the join attribute. Now, \sys has acquired all the values of this join attribute that could potentially produce the final query output. Therefore, it is able to convert the join operation into an \texttt{IN} filter and apply it to the other table. \grammar{Using the samples from the second table, \sys is able to estimate its selectivity.} As a result, \sys can order other filters along with this \texttt{IN} filter to minimize the expected cost. Essentially, \sys might end up running the join operation ahead of the filters, contradicting the traditional database optimizers.

\grammar{Taking Figure~\ref{fig:example-join} as an example, suppose that \sys chooses to transform the join into an \texttt{IN} operation as a filter on $T_2$, i.e., \texttt{``P.team\_name\,\texttt{IN}\,[Lakers, Celtics, Warriors]''}.} Assume that the selectivity of this filter is $0.1$ as shown in Figure~\ref{fig:example-join}-b. This means that approximately $10\%$ of the tuples in the \texttt{Players} table can be joined with the tuples in the \texttt{Teams} table that satisfy $\theta_1$. Then \sys updates $\theta_2$ to $\hat{\theta_2}=$ \texttt{``P.team\_name \texttt{IN} [Lakers, Celtics, Warriors] \texttt{AND} P.age>35''}. Because this newly generated filter has a relatively low selectivity and cost, \sys will prioritize it over existing filters based on the principle discussed in Section~\ref{subsec:filter}. This filter prunes most documents before running other filters (i.e., P.age>35), thus significantly reducing the LLM cost.

Because we have two options to convert the join operation, namely an IN filter either on table $T_1$ or table $T_2$, we establish two cost models respectively, for \sys to make decision.

\noindent [\underline{Plan \ding{173}: \textit{Push $\theta_1$ to $T_1$ and transform the join to filter on $T_2$}}] As discussed above, this plan executes $\theta_1$ on $T_1$ first, and then transforms the join operation to a filter on $T_2$. Its cost is estimated as below:

\vspace{-1em}
\begin{equation}
\label{eq:173}
\small
Cost(\theta_1(T_1) \rightarrow \hat{\theta_2}(T_2)) = 
\sum_{i=1}^{|T_1|} \totalcost_1^i + p_1 \sum_{i=1}^{|T_1|} c_a^i +
\sum_{i=1}^{|T_2|} \hat{\totalcost}_2^i
\end{equation}

\noindent [\underline{Plan \ding{174}:\textit{Push $\theta_2$ to $T_2$ and transform the join to filter on $T_1$}}]This plan executes $\theta_2$ on $T_2$ first, then transforms the join to a filter on $T_1$. Its cost can be estimated in a similar way to that of Plan \ding{173}.

\vspace{-1em}
\begin{equation}
\label{eq:174}
\small
Cost(\theta_2(T_2) \rightarrow \hat{\theta}_1(T_1)) = 
\sum_{i=1}^{|T_2|} \totalcost_2^i + p_2 \sum_{i=1}^{|T_2|} c_{a}^i +
\sum_{i=1}^{|T_1|} \hat{\totalcost}_1^i
\end{equation}

For Plan \ding{173}, the cost on $|T_1|$ remains the same compared to Plan \ding{172}. However, the cost on $T_2$ changes since this plan prioritizes the filter on $a'$. Now the cost is $\scriptstyle\mathit{\sum_{i=1}^{|T_2|} \hat{\totalcost}_2^i=|T_2|\times(c_{a'}^i + p_{a'}\times c_2^i) = 51 \times (30 + 0.1 \times 50) = 1785}$. Here, $p_{a'}$ is the selectivity of the \texttt{IN} filter. Then the overall cost becomes $\scriptstyle\mathit{Cost(\theta_1(T_1) \rightarrow \hat{\theta_2}(T_2)) = 1590 + 1785 = 3375}$.

\noindent\textbf{Selecting a Plan.}
Next, we discuss how to use the above cost models to produce a query plan. First, we present a lemma showing that Plans \ding{172} and \ding{173} are at least as good as Plan \ding{172} in all cases. The proof can be found in our technical report~\cite{full_version}. Then, we show how \sys picks one between Plan \ding{173} and Plan \ding{174}.




\begin{lemma}
    Given a query $Q$ containing a join operation, the expected cost of Plan \ding{172} is always greater than or equal to that of Plan \ding{173} and Plan \ding{174}, i.e., $Cost(\theta_1(T_1) \bowtie \theta_2(T_2)) \geq Cost(\theta_1(T_1) \rightarrow \hat{\theta_2}(T_2))$, and $Cost(\theta_1(T_1) \bowtie \theta_2(T_2)) \geq Cost(\theta_2(T_2) \rightarrow \hat{\theta}_1(T_1))$.
\end{lemma}

\grammar{Next, \sys has to make a decision between Plan \ding{173} and Plan \ding{174}.}
\add{ The simplest approach is to calculate the cost using Equation~\ref{eq:173} and~\ref{eq:174}. While the first two terms are straightforward, the challenge lies in determining the third term, specifically the selectivity of $\theta(\texttt{IN})$. This is essentially about estimating the join selectivity between $T_1$ and $T_2$. Precise estimation of join selectivity is a known issue in databases. Worst yet, in \sys, the optimizer only has access to samples of $T_1$ and $T_2$, not the full tables, before executing the query. }

Nonetheless, we state that it is typically sufficient to decide between Plans \ding{173} and \ding{174} by only considering the first two terms. 
This is because a small sum of these two terms suggests that the number of tuples, either before or after applying the filters, is small.
Consequently, the selectivity of $\theta(\texttt{IN})$ is low, which likely in turn leads to a small third term. Therefore, if a plan has a smaller cost on the first two terms than the other, its overall cost also tends to be smaller. Thus, given $T_1, T_2$, if $\sum_{i=1}^{|T_1|} \totalcost_1^i + p_1 \sum_{i=1}^{|T_1|} c_a^i < \sum_{i=1}^{|T_2|} \totalcost_2^i + p_2 \sum_{i=1}^{|T_2|} c_{a'}^i$, we choose Plan \ding{173}, otherwise we choose Plan \ding{174}.

\noindent\textbf{Query Execution in \sys: Mixing Query Optimization With Execution.}
After selecting a plan ($e.g.$, Plan \ding{173}), \sys extracts the values of the join attribute $T_1.a$, executes the filters on it if there are any, and then transforms the join into an \texttt{IN} filter on $T_2$. Since \sys has obtained all the values of $T_1.a$, it can more accurately estimate the selectivity of \texttt{IN}. \sys then triggers the optimizer again and uses the filter ordering optimization described in Sec.~\ref{subsec:filter} to produce \grammar{the optimal order} to execute the remaining filters. 


\subsubsection{Adaptive Join Ordering\nopunct}\ \\ 
\label{sec.join.adaptive}
Next, we discuss the join ordering strategy, which orders multiple joins involved in a query, a classical yet challenging problem in relational databases. Given a set of tables $\mathcal{T}=\{T_1, T_2, ..., T_{|\mathcal{T}|}\}$,  users could specify a join graph $\mathcal{G}=\{\mathcal{T}, \mathcal{E}\}$, where each edge indicates a join between two tables. In addition, $\theta_i, i\in [1, |\mathcal{T}|]$ denotes the corresponding filters on table $T_i$.

To produce an optimal join order, the classical database optimizers, e.g., the Selinger optimizer~\cite{Selinger_optimizer}, utilize the dynamic programming algorithm~\cite{expression_tree} that depends on some key statistics, such as the cardinalities of the intermediate join results. Estimating these cardinalities becomes rather challenging in our scenario, as the attribute values have not been extracted from unstructured documents yet. 


We introduce an adaptive join ordering approach to discover the optimal join order. First, it chooses one single join by iterating every edge $e\in \mathcal{E}$, i.e., every join in the query, estimating its cost, and choosing the join with the minimal cost. The cost of each join can be directly estimated using the cost model described in Section~\ref{subsec:single}. Then it determines the join plan  (Plan \ding{173} or \ding{174}) of this selected join operation and immediately executes this join. We use $T'$ to denote the result of the first join. Next, it finds another table to join with $T'$, forming a left deep query plan. More specifically, we use $J(T')$ to denote the set of tables that can join with $T'$. \sys selects the one in $J(T')$ that incurs the minimal cost. To this end, \sys has to estimate the cost of $T' \bowtie T_j, T_j\in J(T')$. As discussed in Sec.~\ref{subsec:single}, because $T'$ has already been available, \sys transforms this join to a \texttt{IN} filter operation and estimates its cost, i.e., $\sum_{i=1}^{|T_j|} \hat{\totalcost}_j^i$. Afterward, it joins $T'$ with the selected table according to the join plan. \sys repeats this process until all joins are conducted.

\end{sloppypar}
\section{Index-based Attribute Extraction}
\label{sec:IndexDefinition}

In addition to avoiding unnecessary data extraction operations, we propose an index-based strategy to further reduce the cost of each extraction and improve the accuracy.


\subsection{Two-level Index Construction}\label{sec:indexing}
 
\sys starts with constructing a document-level index. Given a document set $\docs$ as input, \sys first uses the NLTK package \grammar{to generate a document summary efficiently},
\add{which is then transformed into an embedding using
a pretrained model. Here, we choose E5Model~\cite{E5_Wang_Yang_Huang_Binxing_Yang_Jiang_Majumder_Wei_2022} due to its state-of-the-art performance on massive text embedding benchmarks covering diverse retrieval tasks.}

\sys then constructs the segment-level index. For each document $d$ $\in$ $\docs$, \sys dynamically splits the document into relatively small and semantically coherent segments. The goal is to ensure that each attribute can be extracted from a single segment. 
To achieve this, we employ the \texttt{SemanticChunker} function in LangChain to segment text by examining both its syntactic structure and semantic coherence.
%
\add{Initially, the document is divided into sentences. \texttt{SemanticChunker} then evaluates the embedding similarity of consecutive sentences. If sentences are semantically coherent, they merge; if not, they remain separate. This iterative procedure continues for all sentences, enhancing attribute extraction by maintaining semantic coherence.}

%
Eventually, these segments are represented as a set $S$, where each $s \in \mathcal{S}$ is a coherent, self-contained portion of the document's content.
\sys then embeds each segment $s \in \mathcal{S}$ using E5Model. 
Finally, \sys loads document and segment embeddings into two high-dimensional vector indexes for efficient data retrieval. 

\subsection{Searching the Index}\label{sec:search index}

Given a query, \sys first uses the document-level index to search the relevant documents and then uses the segment-level index to identify the relevant segments from the returned documents. 
\noindent \textbf{Document Retrieval.} Once receiving a query, \sys retrieves from $\docs$ the documents that potentially contain an attribute in the query.
To achieve this, \sys first converts attribute names and their descriptions into embeddings. Then it averages these embeddings to generate a final embedding $e(Q)$.
Afterward, based on the document-level index $\mathcal{I}_{\docs}$, \sys searches the documents with an embedding close to $e(Q)$, i.e., $D_Q = \{ d_i|d_i\in \docs, dist(e(d_i), e(Q))<\tau \}$, where $dist()$ is the distance function.
%
\add{While cosine similarity is commonly used for embedding comparison, it is monotonically related to Euclidean distance when vectors are L2-normalized, i.e., \( \|v_1 - v_2\|^2 = 2 - 2 \cdot \cos(v_1, v_2) \). Thus, minimizing normalized Euclidean distance is equivalent to maximizing cosine similarity for ranking. We adopt it here because it is natively supported in vector indexing libraries such as PQ~\cite{product_quantization} and is computationally efficient.}
$\tau$ is a threshold initially set as a high value to guarantee a high recall. However, this may return some irrelevant documents. To solve this problem, \sys automatically adjusts this threshold in the next segment retrieval phase based on the sampled documents to obtain a more precise document set $D_Q^*$.


\noindent\textbf{Evidence Augmented Segment Retrieval.}
As discussed in Section~\ref{subsec:index-based tuple extraction}, to achieve accurate segment retrieval, we propose to sample a small subset (approximately 5\% of $D_Q$) of documents that will be carefully analyzed by an LLM.
To be specific, for the attributes in the attribute set $A_Q$ of query $Q$, \sys asks the LLM to return their values and the segments from which these values are extracted.

Next, we transform these segments into embeddings. The key observation here is that the segments containing the same attribute tend to show some common patterns. For example, in Stephen Curry's profile, the segment about the ``age'' attribute includes "Wardell Stephen Curry II (born March 14, 1988) is an American professional basketball player ...", while the corresponding segment in Kevin Durant's profile mentions that "Kevin Wayne Durant (born September 29, 1988), also known by KD, is an American professional basketball player ...". The patterns of these segments are remarkably similar. Using these patterns as additional evidence could enhance retrieval, addressing the issue where simply using the query embedding may miss attribute-relevant segments, as discussed in Sec.~\ref{subsec:index-based tuple extraction}.

\add{When no relevant segments are found for an attribute \(a_i\) in the sampled documents, \sys leverages the LLM to synthesize evidence. It prompts the LLM with the attribute name, its description, and optional contextual information to generate a small number (e.g., 20) of representative text segments, which are then embedded.}

%
\add{However, using all these embeddings as evidence tends to unnecessarily introduce redundancy and, in turn, produce too many candidate segments.} We thus propose to utilize the $k-$means algorithm to group these embeddings (with a relatively small $k$, such as 3) and only use the cluster centers as evidence. 
We use $\overline{e}^j_i, j\in[1,k]$ to denote one of the cluster centers, i.e., one piece of evidence for $a_i$ ($k$ pieces in total for each attribute).

%


\noindent\underline{\textit{Segment Retrieval.}}
Next, we discuss how to use the collected evidence to retrieve the segments relevant to an attribute in the query, $e.g.,$ the attributes \texttt{age} and \texttt{all-stars} in the query $Q$ shown in the middle block of Figure~\ref{fig:framework}.
%
For each document in $D_Q^*$, \sys utilizes the segment-level index $\mathcal{I}_S$ to find the relevant segments for a given attribute $a_i$ based on each piece of evidence $\overline{e}^j_i$. Next, \sys combines the retrieved segments in each document with respect to the evidence of $a_i$, removes the duplicate segments, and feeds them into the LLM for attribute extraction.

\noindent\underline{\textit{Setting the Threshold.}}
\sys uses two distance thresholds to determine whether a document or a segment is relevant to an attribute. Setting these thresholds appropriately is critical to ensure the quality of retrieval. Setting these distance thresholds too high tends to return many irrelevant segments and hence increase extraction cost, while setting them too small might miss relevant segments, in turn impacting the extraction quality. Clearly, relying on users to manually set these thresholds correctly is challenging. \sys thus introduces an automatic thresholding method to solve this problem.  

For the threshold $\tau$ used to find documents relevant to a query $Q$, \sys initially sets it as a high value to avoid missing relevant documents. \sys then adjusts it to an appropriate value by analyzing the LLM extraction results on the sampled documents. 
Specifically, \sys first uses the high $\tau$ to obtain a set of documents denoted as $D_Q$. \sys then samples a subset $D_Q^S$ from $D_Q$ and uses LLM to extract the attributes of a table from $D_Q^S$. Based on the result of the extraction, \sys divides $D_Q^S$ into two subsets: $D_Q^n$, which consists of documents lacking attribute information therefore deemed irrelevant to the query, and $D_Q^m = D_Q^S \setminus D_Q^n$, which contains relevant documents. The maximum Euclidean distance between the embeddings of documents in $D_Q^m$ and the embedding $e(Q)$ serves as the threshold, i.e., $\small \tau = \max \{dist(e(d_i), e(Q)) | d_i \in D_Q^m\}$. Intuitively, this new $\tau$ threshold will exclude irrelevant documents.

Similarly, \sys leverages the sampled documents to set the threshold $\gamma_i$ to retrieve the segments containing an attribute $\mathit{a_i}$, $\mathit{i\in[1,M]}$. More specifically, $\gamma_i$ is set as the maximal distance between any pair of segments containing the value of $a_i$, i.e., $\small\gamma_i = \max \{ dist(E_i[x], E_i[y]) | \forall x,y \in [1, |E_i|], x \neq y \}$,
where $E_i$ represents the set of segments related to attribute $a_i$ in $D_Q^m$. $\mathit{E_i[x]}$ and $\mathit{E_i[y]}$ denote the $x$-th and $y$-th embeddings of two segments in $E_i$. To be cautious, in implementation we increase $\gamma_i$ by 0.1, that is, $\gamma_i  = \gamma_i + 0.1$; and equally we adjust $\tau$.


\section{Experimental Evaluation}
\label{sec: eval}

\subsection{Experimental Settings}\label{exp: setting}

\begin{table}
    \centering
    \resizebox{0.4\textwidth}{!}{
    \begin{tabular}{|c|c|c|c|c|} \hline 
         &  Avg. \#-Tokens&  \#-Doc& \#-Attributes& \#-Quries \\ \hline 
         \legal &  6247&  100 & 10& 10\\ \hline
         \wiki &  1264&  200 & 20& 25\\\hline  
         \swde &  416&  200 & 16& 15 \\\hline
    \end{tabular}}
    \caption{Datasets}
    \vspace{-2em}
    \label{tab:datasets}
\end{table}



\noindent \textbf{Datasets.}
We use 3 datasets with 500 documents in total, covering diverse domains, and 50 queries in various types. We employ human evaluators verifying the attributes extracted by LLMs from documents, and thus establish the ground truth.


\noindent \underline{\legal}~\cite{lcr} includes 3,000 case reports. We sample 100 documents from them. Each document averages 6,247 tokens and contains detailed information such as the court, judge, and legal reasoning.

\noindent \underline{\wiki.}
We crawl 200 Wikipedia pages across 10 domains,  such as directors, cities, NBA players, companies, etc., some of which can be joined; the average number of tokens per document is 1,264.
%

\noindent \underline{\swde}~\cite{SWDE_Hao_Cai_Pang_Zhang_2011} is a dataset used in our baseline~\cite{T2T_arora2023evaporate}. We sample 200 web pages, with each averaging 416 tokens. Despite the relatively short length of the documents, \swde contains 16 attributes.

The datasets vary in length, structure, and domain: \wiki has a hierarchical structure across various domains, \swde includes short documents, and \legal features long documents from a single domain. They facilitate thorough evaluations in various scenarios.

\noindent\underline{\textit{Ground Truth Generation.}}
We organize each dataset into domains of similar documents, sample 5 documents per domain, and use LLMs to identify key attributes, as shown in Table~\ref{tab:datasets}. We utilize LLMs for attribute extraction from all documents, verified by 10 graduate students.

\noindent \underline{\textit{Query Construction.}}
We create queries for single tables and join tables. The queries cover both range and equality filters. We first construct the filters in \texttt{WHERE} clause: 
(1) For each query, we randomly sample a certain number of attributes from the attribute set in the query to construct the filters; (2) For numerical attributes, we randomly create different types of filters, including $=$, $\leq$, and $\geq$, while for categorical attributes, we only generate equality filters; (3) We then use these single filters to construct conjunctions, disjunctions, or a combination of both. Each of these three categories has roughly the same number of queries. Next, we randomly sample a certain number of attributes to form the \texttt{SELECT} clause. Finally, we ask graduate students to validate all queries and eliminate the unreasonable ones.


\noindent\textbf{Baselines.} We compare \sys with various baselines.
%

\noindent (1) \zendb\cite{T2T_lin2024zendb} adopts a hierarchical semantic tree to extract tuples from documents and uses SQL-like queries for analysis.

\noindent (2) \pz (Palimpzest)~\cite{MMDB_liu2024pz} allows users to convert and analyze unstructured data with a declarative language hosted in Python. The existing PZ prototype offers basic optimizations on the usage of LLMs, while in-depth optimizations are still ongoing.

\noindent (3) \lotus\cite{MMDB_patel2024lotus} supports a bunch of LLM-powered operations to analyze documents. It features some basic optimizations to improve the accuracy and query latency.


\noindent (4) \rag~\cite{rag_baseline}  embeds attributes and their descriptions for similarity search. In contrast to \sys, it does not incorporate a document-level index and does not utilize evidence to enhance retrieval. 


\noindent (5) \nlp\cite{closedIE_1} uses a model that has been fine-tuned using a vast quantity of labeled (attribute, value) pairs to extract relevant information from a given context in response to a query.

\noindent (6) \evaporate (Evaporate)\cite{T2T_arora2023evaporate} is an LLM-based data extraction method. Rather than routinely employing LLMs to extract values from each document, it instead uses LLMs to automatically synthesize code. 


\noindent (7) \sys is our full-fledged solution.


\noindent\textbf{Evaluation Metrics.}
We measure accuracy, cost, and latency with respect to all queries. 
%
%
For accuracy, we evaluate the average precision, recall, and F1-score across all queries. 
Given a query $Q$, the set of tuples returned by a method is denoted as $T(Q)$, and the ground truth is denoted as $GT(Q)$. A tuple $t\in T(Q)$ is considered correctly extracted only if all its cell values match the corresponding ground truth values. Therefore, we have $P = \frac{|T(Q)\cap GT(Q)|}{|T(Q)|}$, $R = \frac{|T(Q)\cap GT(Q)|}{|GT(Q)|}$ and $F1 = \frac{2*P*R}{P+R}$.
\add{For LLM cost, we measure the average number of input and output tokens per document for each query, including the cost of sampling, to reflect the full processing of \sys.}
For latency, we measure the mean execution time of queries per document.

\subsection{Comparison with Baselines} \label{sec:exp:baselines}

\begin{table*}[h]
  \centering
        \scriptsize
        \resizebox{0.7\linewidth}{!}{%
        \begin{tabular}{|c|c|c|c|c|c|c|c|c|c|}
        \hline
        \multirow{2}{*}{} & \multirow{2}{*}{}  & \multicolumn{1}{c|}{\nlp} & \multicolumn{1}{c|}{\evaporate} & \multicolumn{1}{c|}{\rag} & \multicolumn{1}{c|}{\pz} & \multicolumn{1}{c|}{\zendb} & \multicolumn{1}{c|}{\lotus} & \multicolumn{1}{c|}{\sys} \\ \hline
        \multirow{3}{*}{\textbf{\texttt{Wiki}}} 
      & Precision  &0.33 &0.39 &0.79&0.75&0.81&\textbf{0.93}&\textbf{0.93} \\ 
        & Recall  &0.19 &0.29 &0.24&0.81&0.78&\textbf{0.84}&0.82\\
        & F1-score  &0.24 &0.33 &0.37&0.78&0.79&\textbf{0.90}&0.87 \\ 
        \hline
        \multirow{3}{*}{\textbf{\swde}} 
        & Precision  &0.72 &0.85 &0.88&0.90&0.82&\textbf{0.95}&0.94 \\ 
        & Recall  &0.51 &0.59 &0.78&0.83&0.86&\textbf{0.98}&0.97\\ 
        & F1-score  &0.59 &0.70 &0.83&0.86&0.84&\textbf{0.96}&0.95\\ 
        \hline
        \multirow{3}{*}{\textbf{\legal}} 
        & Precision  &0.11 &0.16 &0.42&0.43&0.45&0.45&\textbf{0.63}\\ 
        & Recall  &0.07 &0.13 &0.29&0.61&0.73&0.45&\textbf{0.84}\\ 
        & F1-score  &0.09 &0.14 &0.34&0.50&0.55&0.45&\textbf{0.72}\\ 
        \hline
        \end{tabular}%
        }
        \caption{Accuracy Comparison}
        \vspace{-3.7em}
        \label{tab:baselines}
\end{table*}

\begin{table}[h]
    \centering
        \resizebox{0.85\linewidth}{!}{
        \begin{tabular}{|c|c|c|c|c|c|c|c|}
        \hline
        \multirow{2}{*}{\textbf{Method}} & \multicolumn{3}{c|}{\textbf{\#-Token Cost}} & \multicolumn{3}{c|}{\textbf{Latency (s)}} \\ \cline{2-7} 
         & \textbf{\texttt{Wiki}} & \textbf{\swde} & \textbf{\legal} & \textbf{\texttt{Wiki}} & \textbf{\swde} & \textbf{\legal} \\ \hline
          \evaporate & - & - & - &\textbf{0.06}  &\textbf{0.05}  &\textbf{0.08} \\ 
         \nlp & - & - & -  &0.76  &0.65  &1.83 \\ 
         \zendb & 260 & 280 & 2530 & 2.08 & 1.77 & 2.73 \\
         \pz & 400 & 320 & 2610 & 2.17 & 2.16 & 2.85 \\ 
        \rag & 440 & 340 & 3500 & 2.57 & 2.65 & 3.01 \\ 
        \lotus & 2520 & 1150 & 12480 & 2.66 & 2.78 & 3.36 \\ 
        \sys & \textbf{170} & \textbf{190} & \textbf{2030} & 1.12 & 1.21 & 2.68 \\ \hline
        \end{tabular}
        }
        \caption{Cost and Latency Comparison}
        \vspace{-2em}
        \label{tab:costlatency}
\end{table}

\noindent \textbf{Accuracy w.r.t. All Queries.} By the results (including $P$, $R$ and $F1$) shown in Table~\ref{tab:baselines}, these method are ranked as follows: \sys $\approx$ \lotus $>$ \pz $>$ \zendb $>$ \rag $>$ \evaporate $>$ \nlp.

\sys and \lotus achieve the highest accuracy among all baselines. \lotus performs well because it uses an LLM to scan every piece of text of all documents, leading to extremely high LLM costs.  \sys is competitive with \lotus but with a much lower cost, because our two-level index and the evidence-augmented retrieval method accurately identify relevant segments for queries, feeding only these to LLMs. \sys thus ensures quality with low LLM cost. On \legal dataset, \sys achieves an F1-score of 0.7, much higher than that of \lotus (0.45). The reason is that the documents in \legal contain a large number of tokens, including much irrelevant information that misleads LLMs and causes hallucinations.

\sys outperforms \zendb because: (1) \sys leverages the evidence to identify relevant segments, which is more accurate than \zendb that simply uses the attribute description to find the most relevant sentence; (2) although \zendb uses a semantic tree to locate each attribute, in practice, many documents, e.g., the documents in the \legal dataset, are not well-structured, making it hard for \zendb to construct an effective tree. Therefore, the F1-score of \zendb on \legal is only 0.52, much lower than that of \sys (0.7). Similarly, \sys outperforms \pz because its existing prototype lacks an effective text segment retrieval component.

\begin{figure}[h!]
    \centering
    \includegraphics[width=1\linewidth]{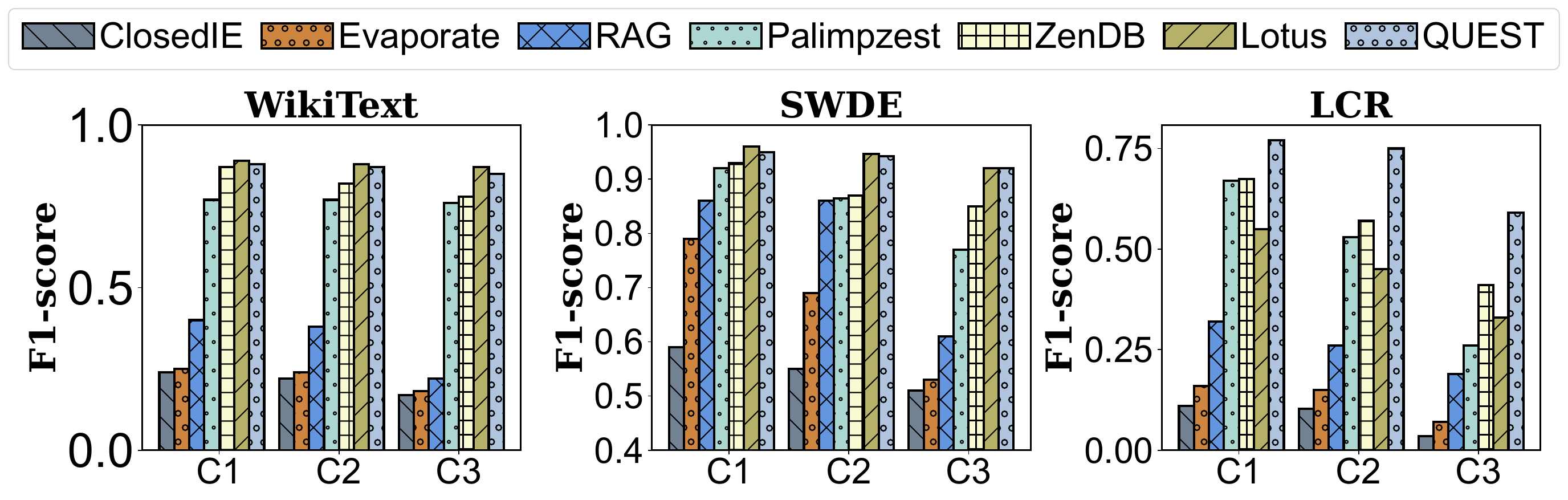}
    \vspace{-2.5em}
    \caption{F1-Score of Baselines (Different Query Groups)}
    
    \label{fig:main f1}
\end{figure}

\begin{figure}[h!]
    \centering
    \includegraphics[width=1\linewidth]{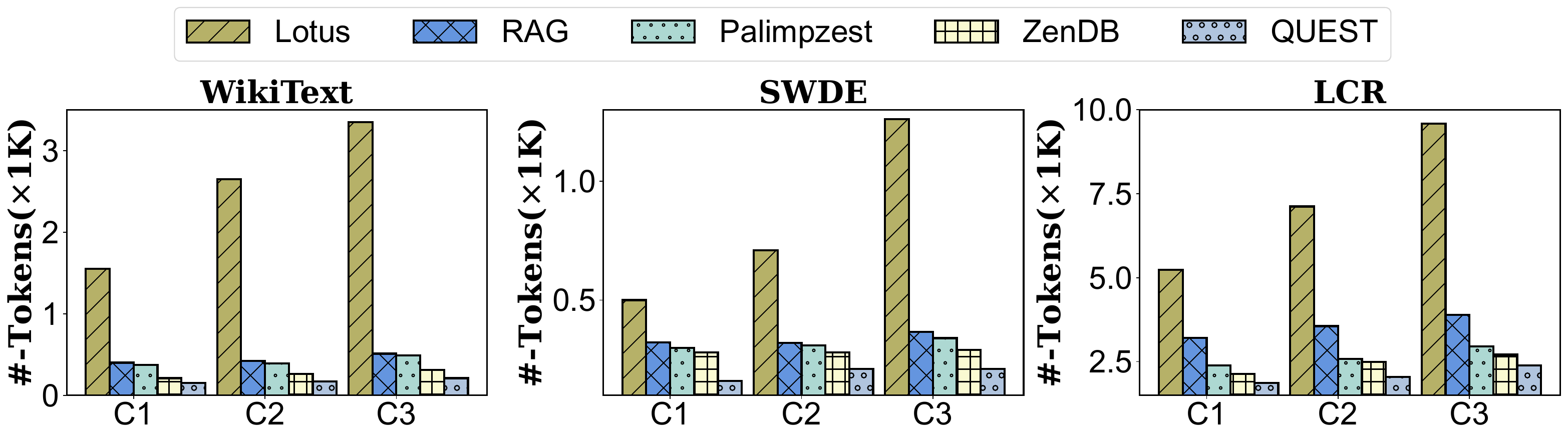}
    \vspace{-2.5em}
    \caption{Cost of Baselines (Different Query Groups)}
    \vspace{-0.7em}
    \label{fig:main cost}
\end{figure}

As expected, the \rag-based method is less effective than \sys because the embedding of attribute and its descriptions is often not informative enough, thus tending to miss segments that are highly relevant to the query but do not possess similar phrases. Our evidence augmented retrieval strategy successfully solves this issue. 
Evaporate uses LLMs to generate code for data extraction, aiming to reduce LLMs costs. However, it does not perform well on accuracy because code essentially corresponds to a limited number of rules, which tend to be less effective when handling complex documents. 
\nlp performs poorly as pre-trained NLP models lack generalizability across domains.

\noindent \textbf{Overall Cost.}
Table~\ref{tab:costlatency} shows the LLM cost of the baselines that have reasonable accuracy. We do not report the cost of \nlp and \evaporate. This is because the first one does not use LLM, while \evaporate just spends a few tokes on generating code. However, the accuracy of these two methods is rather low. The other methods are ranked as follows: \sys $<$ \zendb $<$ \pz $<$ \rag  $<$ \lotus. \sys is the most cost-effective because (1) it has a two-level index that can precisely locate a small number of segments where an attribute can be extracted, and (2) the filter reorder optimization minimizes the frequency of its LLM invocations by early terminating \grammar{the evaluation of the filters} whenever possible. 

 \lotus is the most expensive method because it feeds the entire document to LLMs for each filter. 
For example, on dataset \legal, \lotus costs 6$\times$ more tokens than \sys. \rag is cheaper than \lotus because it feeds only a subset of segments to LLMs rather than the entire document.
\pz and \zendb save more cost than \rag because they reorder filters based on selectivities. However, both methods are more costly than \sys. For example, on dataset \wiki, \pz and \rag consume more than 2$\times$ tokens than \sys because they process every document in the same order, while \sys generates the optimal order per document.

\noindent \textbf{Overall Latency.} 
Table~\ref{tab:costlatency} shows the query latency.
\nlp and  \evaporate are the fastest, although their accuracy in general is low. This is because \nlp does not use LLMs, while \evaporate only spends a few tokens on code generation. 
For other LLM-based methods, \sys is about 2$\times$ faster than \pz, \rag, \lotus and \zendb.
This is because, in these methods, the LLM inferences dominate the query execution time, while the fine-grained filter ordering strategy of \sys reduces both the number of LLM calls and the number of tokens consumed per call.
\lotus is the slowest because it has to send each document to the LLM when evaluating a filter.

\noindent \textbf{Varying the Number of Filters.} We evaluate \sys's performance with varying filter numbers, categorizing queries into: $\texttt{C1}$ with one filter, $\texttt{C2}$ with 2-3 filters, and $\texttt{C3}$ with 4 or more. We can observe in Figure ~\ref{fig:main f1} that as the number of filters grows, the accuracy of all baselines decreases because more filters tend to introduce more errors during attribute extraction. The cost of almost all baselines 
increases because more filters invoke more LLM calls. However, the increase of \sys is the slowest, thanks to our filter ordering that reduces unnecessary attribute extraction.

\subsection{Comparison of Filter Ordering Strategies}
\begin{figure}[h!]
\vspace{-0.6em}
    \centering
    \includegraphics[width=0.9\linewidth]{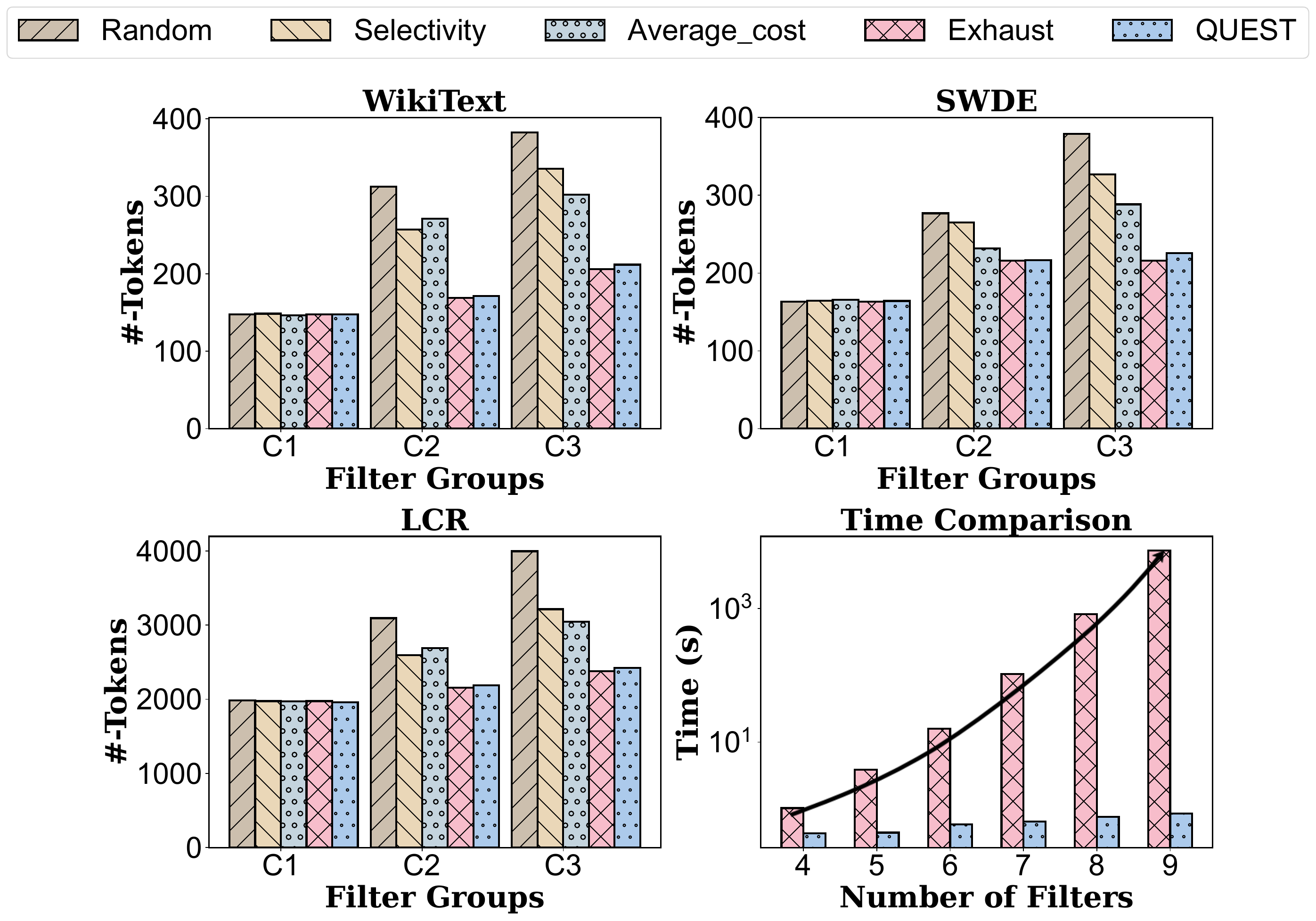}
    \vspace{-1.5em}
    \caption{Comparison of Filter Reordering Strategies}
    \vspace{-0.7em}
    \label{fig:order}
\end{figure}
We evaluate the following baselines: (1) \texttt{Random:} The filters are executed in random order; (2) \texttt{Selectivity:} The filters are ordered based on the selectivity; (3) \texttt{Average\_cost:} The filters are ordered based on both the selectivity and the estimated average cost of extracting each attribute from the sampled documents; (4) \texttt{Exhaust:} It exhaustively enumerates all possible orders and returns the optimal one per document.

In Figure~\ref{fig:order}, for queries in \texttt{C1}, the cost of all baselines is almost identical because there is only one filter and hence one order per query. For queries with more filters, these methods are ranked as follows by the LLM cost: \sys $\approx$ \texttt{Exhaust}  $<$ \texttt{Average\_cost} $<$ \texttt{Selectivity} $<$ \texttt{Random}. This shows that \sys produces optimal orders in most cases. As the number of filters increases, \sys's performance improves due to increased optimization opportunities.

Furthermore, we evaluate the scalability of \sys and \texttt{Exhaust} when handling queries with a relatively large number of filters. Figure~\ref{fig:order} shows that as the number of filters increases, the run time of \texttt{Exhaust} explodes due to the exponential time complexity. Conversely, the run time of \sys increases slowly, indicating that it is more efficient and effective for complex queries. 



\subsection{Evaluation of Join}
\begin{figure}[h!]
\vspace{-0.8em}
    \centering
    \includegraphics[width=1\linewidth]{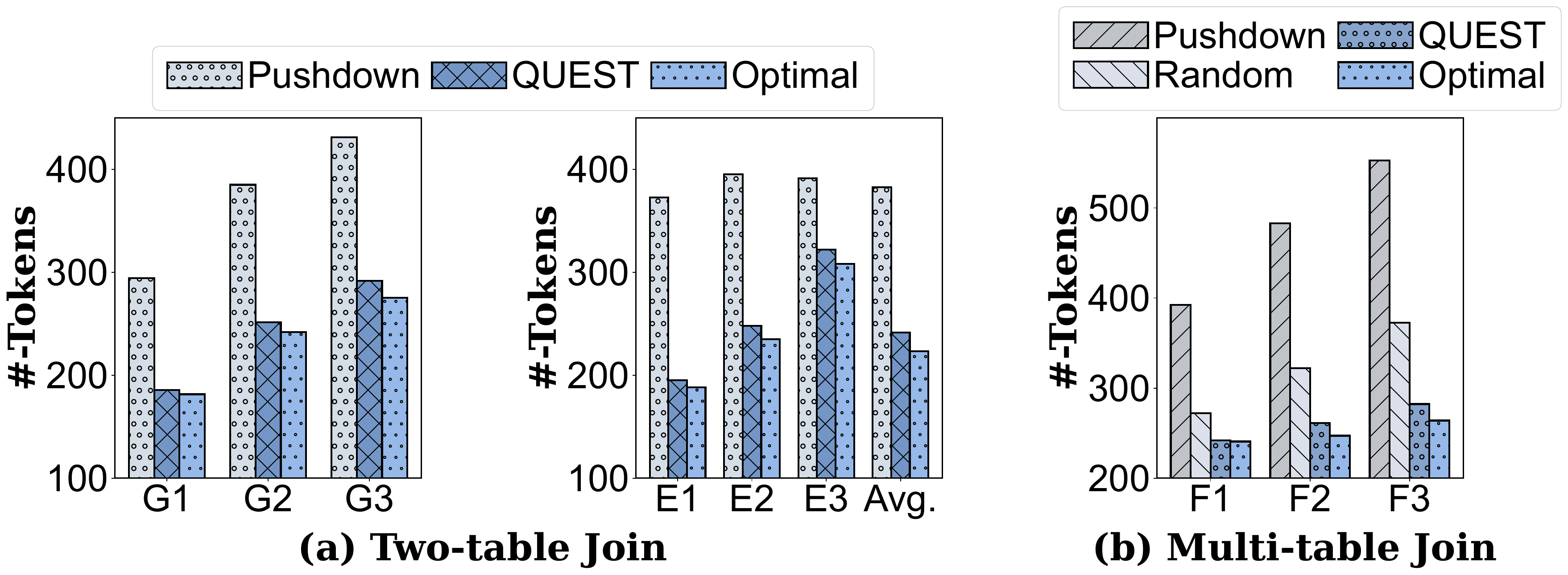}
    \vspace{-2.5em}
    \caption{Evaluation of Join}
    \vspace{-0.5em}
    \label{fig:table join}
\end{figure}

\noindent \textbf{Tables to be Joined.} We define 4 tables (i.e., \texttt{Player}, \texttt{Team}, \texttt{City}, \texttt{Owner}) in \wiki dataset. \texttt{Player} and \texttt{Team} join on the \texttt{team\_name} attribute. \texttt{Team} and \texttt{City}
join on the \texttt{location} attribute. \texttt{Team} and \texttt{Owner}
join on the \texttt{owner\_name} attribute. We construct filters for each table using the method in Section~\ref{exp: setting}.

\noindent \textbf{Two-table Join.} To demonstrate effectiveness, we compare \sys with the typical optimization in relational databases, so-called \texttt{Pushdown}, which always pushes down the filters before performing the joins. We also compare \sys with the \texttt{Optimal} plan, which is obtained by assuming the selectivity of each filter is known.


We create queries incorporating one of the three aforementioned joins, along with some random filters. In total, we construct 60 queries. We classify these queries into three categories based on the number of filters, as previously stated. We execute the set of queries (\texttt{G1}-\texttt{G3}) and present the mean token consumption. In Figure~\ref{fig:table join}-a, \sys significantly reduces cost compared to \texttt{Pushdown} because it transforms a join to a filter operation and effectively orders the filters to minimize the frequency of LLM invocations. In this way, \sys has the opportunity to run a join first if it incurs a small data extraction cost. In particular, \sys only costs slightly more than \texttt{Optimal} due to its effective optimization strategies.

Next, we record the selectivity of each \texttt{IN} filter according to the plan chosen by \sys for each query. We then classify the queries into three new categories ($E_1-E_3$) based on the selectivities. The first group, $E_1$, corresponds to the selectivities ranging from $0-0.3$, while $E_2$ and $E_3$ correspond to $0.3-0.6$ and $0.6-1$, respectively. Moreover, we present the average cost associated with all queries. Figure~\ref{fig:table join}-a illustrates that when the selectivity of the \texttt{IN} filter is low, it tends to be executed first, leading to lower cost than the traditional predicate pushdown strategy. Conversely, as the selectivity rises, the \texttt{IN} filter might fall back to the traditional predicate pushdown strategy, thus always finding a plan \grammar{with the lowest cost.}

\begin{figure}[h!]
\vspace{-0.5em}
    \centering
    \includegraphics[width=\linewidth]{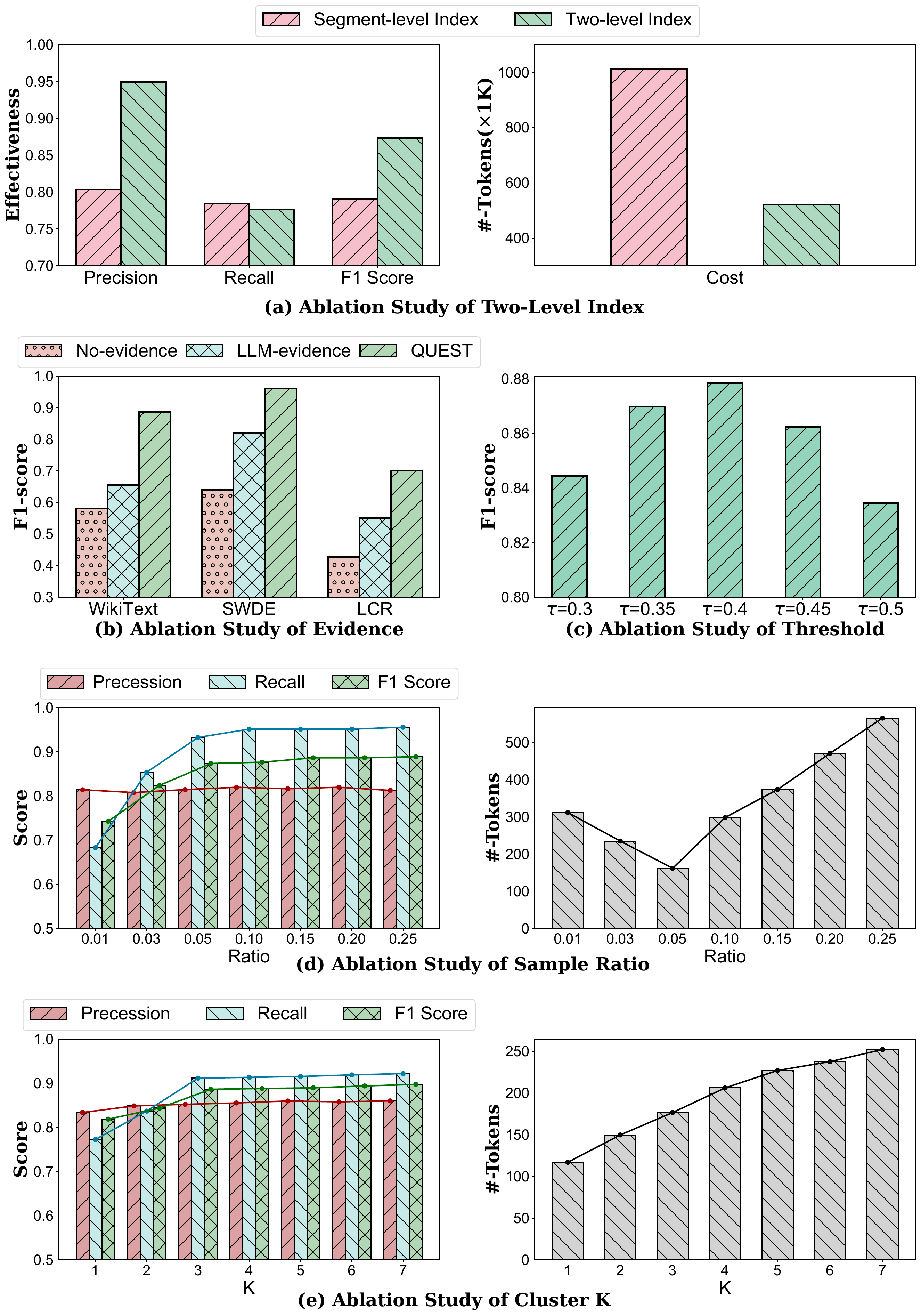}
    \vspace{-2.5em}
    \caption{Ablation Studies}
    \vspace{-0.7em}
    \label{fig:ablation}
\end{figure}

\noindent \textbf{Multi-table Join.} Next, we evaluate \sys on queries with multiple joins. Similarly, we create three groups of queries by the number of filters(\texttt{F1}-\texttt{F3}), and each group has 10 queries. The queries in each group apply the same filter operations, while queries in different groups apply different joins. To evaluate the effectiveness of \sys, we compare it with (1) \texttt{Random}: each time we randomly select two tables to join using the above technique; (2) \texttt{Pushdown}: we push down all filters first and then run the joins and (3) \texttt{Optimal}: we assume that the selectivity of each filter is known and enumerate all possible join orders to obtain the optimal one.

In Figure~\ref{fig:table join}-b, the system outperforms \texttt{Random} and \texttt{Pushdown},  comparable to the optimal plan in LLM cost. It outperforms \texttt{Random} because during query execution, \sys dynamically selects the join operation that leads to the lowest cost.
\sys outperforms \texttt{Pushdown} since in unstructured document analysis, pushing down filters first does not always improve the efficiency of join queries.


\subsection{Ablation Studies}
\label{sec:ablation study}

\noindent \textbf{Ablation Study of Two-Level Index.}
To demonstrate the effectiveness of our index, we conduct ablation studies on \wiki, comparing with the baseline that only uses the segment-level index. Here, we record the total cost incurred by each query across all documents and calculate the average over the queries.
In Figure~\ref{fig:ablation}-a, the two-level index achieves a higher F1-score and a lower cost compared to the baseline because the segment-level index selects irrelevant documents, decreasing precision and increasing costs due to unnecessary document processing. 


\noindent \textbf{Ablation Study of Evidence.}
\add{We evaluate \sys against two baselines: \texttt{No-evidence}, relying solely on the attribute and description, and \texttt{LLM-evidence}, using LLM-generated synthetic text to enhance the query. Figure~\ref{fig:ablation}-b shows \sys surpasses both in accuracy by utilizing document-based evidence for enhanced structure and semantic reflection, leading to more precise retrieval.
}

\noindent \textbf{Ablation Study of $\tau$.}
We evaluate the effectiveness of our strategy of setting the threshold $\tau$ automatically. Given a query about NBA players, our adaptive strategy sets $\tau=0.4$. Then, we vary $\tau$ around 0.4.
In Figure \ref{fig:ablation}-c, if $\tau$ is large, the accuracy decreases because more irrelevant documents are retrieved. 
If $\tau$ is small, the accuracy also decreases because of missing many relevant documents. 

\add{\noindent \textbf{Ablation Study of Sample Rate.} 
We perform ablation studies on the \wiki dataset, adjusting the sampling rate near the default 5\%. Figure~\ref{fig:ablation}-d shows that accuracy initially rises with more samples but levels off quickly. Costs decrease first due to better selectivity estimation, then increase with excessive sampling due to LLM overhead. Overall, 5\% efficiently balances quality and cost.}

\add{\noindent \textbf{Ablation Study of Cluster K.} 
We examine the \wiki dataset by varying the cluster count \(K\) near the default of 3. Figure~\ref{fig:ablation}-e demonstrates that while accuracy rises with additional clusters providing richer evidence, it soon levels off due to limited extra information. Precision remains stable, depending largely on extraction strategies. Costs rise with \(K\) as more clusters bring more evidence vectors and retrieved segments, increasing token usage. }

\section{Related Work}


\noindent \textbf{Language Models for Multi-Modal Data Analysis.} Many works focus on analyzing various types of data from diverse sources.
Lotus~\cite{MMDB_patel2024lotus} introduces several semantic operators to facilitate bulk semantic processing, including searching, extraction, and indexing, which can be used to build complex pipelines. However, it routinely utilizes LLMs to analyze the full text, incurring significant LLM costs. Some other works~\cite{MMDB_Chen_Gu_Cao_Fan_Madden_Tang,MMDB_jo2024thalamusdb,MMDB_Thorne_Yazdani_Saeidi_Silvestri_Riedel_Halevy_2021,MMDB_Urban_Binnig}
also apply LLMs or pre-trained language models to analyze unstructured documents. \grammar{Like Lotus, they do not focus on optimizing the language model cost}.

Palimpzest~\cite{MMDB_liu2024pz} analyzes unstructured data with a declarative language. Its optimizer produces an execution plan that uses LLMs to extract and analyze the data. 
However, its optimizations mainly focus on choosing suitable LLMs for tasks, code synthesis, or prompting strategies, which, although effective, are unrelated to classical query optimization principles such as filter ordering or join optimization.
%
CAESURA~\cite{MMDB_urban2023caesura} uses LLMs for natural language analysis over multi-modal data, decomposing queries into operators and invoking models like VisualQA (image-based) and TextQA (text-based) for question answering on different multi-modal tasks, without optimizing LLM costs.
%
UQE~\cite{UQE_dai2024uqe} leverages LLMs for SQL-like analysis over multi-modal data, primarily enhancing aggregation queries with a sampling technique without considering filter ordering or join optimization.

\noindent \textbf{Retrieval Augmented Generation.} RAG~\cite{RAG_Cai_Wang_Liu_Shi_2022,RAG_fan2024survey,RAG_gao2023retrieval,RAG_Li_Su_Cai_Wang_Liu,RAG_Sarthi_Abdullah_Tuli_Khanna_Goldie_Manning_2024} is a popular method for tasks like question answering and is well-suited for attribute extraction, a key operation in \sys, because it retrieves relevant segments using an index. However, Sec.~\ref{sec:exp:baselines} shows that the typical RAG's generic index-based retrieval is less effective than our customized two-level index and evidence augmented strategy.

\noindent \textbf{Text-to-Table Extraction.} Some works focus on extracting structured information from unstructured data. ZenDB~\cite{T2T_lin2024zendb} constructs a semantic hierarchical tree within each document to identify the sections that potentially contain a target attribute. Then, a single matching sentence, as well as several summaries within a section, are fed into an LLM for data extraction.
It also uses several optimizations, including filter ordering and predicate pushdown. However, ZenDB lacks fine-grained document-level query optimization and its index heavily relies on templated document structures. Unfortunately, in practice, abundant of documents lack a clear hierarchical structure. In particular, for documents with lengthy paragraphs, it is rather difficult to simply rely on several summaries and one sentence that matches the text of a query to identify the target attribute.
EVAPORATE \cite{T2T_arora2023evaporate} uses LLMs to extract tables from HTML and PDF files through LLM-based code generation and adopts weak supervision to combine extraction functions. This aims to balance cost and quality. However, relying solely on LLM-generated code for complex documents is not highly accurate, as shown in Sec.~\ref{sec:exp:baselines}.

Another line of work focuses on training models. Closed information extraction~\cite{closedIE_1,closedIE_2,closedIE_3} uses language models to extract query-relevant information from context. Wu et al. \cite{T2T_Wu_Zhang_Li_2022} define text-to-table conversion as a sequence-to-sequence task, improving a pretrained model. Pietruszka et al. \cite{T2T_pietruszka2022stable} uses a permutation-based decoder for text-to-table models, enhancing tasks such as entity extraction. Jiao et al. \cite{T2T_jiao2023instruct} finetune a pretrained model for instruction-following in text-to-table tasks. It extracts structured data, but lacks LLMs' accuracy and cross-domain generalization~\cite{T2T_arora2023evaporate}.

\noindent \textbf{LLMs for Data Preparation.} 
Data preparation is considered a data processing pipeline that converts raw data into an analyzable format, including tasks like data extraction, discovery, cleaning, integration and labeling. Recently, LLMs have shown great skill in data preparation.
Several works finetune LLMs to enhance their abilities in conducting data preparation~\cite{jellyfish_zhang-etal-2024-jellyfish, tablegpt_li2023table}.
%
%
Other works utilize LLMs to address specific tasks, such as data extraction~\cite{T2T_arora2023evaporate}, schema matching~\cite{magneto_liu2024magnetocombiningsmalllarge, llmschema_parciak2024schemamatchinglargelanguage}, data cleaning (including data imputation~\cite{imputation1_ding2024data, imputation2_hayat2024claim}, entity resolution~\cite{ER_fan2024cost}, etc), and data labeling ~\cite{labeling1_bansal2023large, labeling2_xiao2023freeal}.

\section{Conclusion \& Future Work}\label{sec: conclusion}
We propose \sys, a cost-effective LLM-powered system that features novel query optimizations to support unstructured document analysis. By introducing a two-level index, an evidence augmented retrieval strategy, and instance-optimized query execution, \sys effectively reduces the LLM cost while maintaining high accuracy.  Our comprehensive experiments showcase the efficacy of \sys, achieving 30\%-6$\times$ cost saving, while improving F1-score much.



\add{A key future research direction is extending \sys to support aggregation queries. Optimizations such as approximate query processing could estimate the aggregation results by analyzing only a subset of sampled documents. Moreover, we could judiciously create summaries for different possible attributes and directly produce aggregation results from the summaries.
Exploring such strategies would further enhance \sys's capability for comprehensively analyzing unstructured documents.
}



\bibliographystyle{ACM-Reference-Format}
\bibliography{sample_base}

\end{document}